\documentclass[11pt]{article}
\usepackage{jheppub,relsize}
\usepackage{varioref,exscale,latexsym,amsmath,amssymb}
\usepackage{graphicx}
\usepackage{cancel}
\usepackage{inputenc}
\usepackage{amssymb}
\usepackage{natbib}
\usepackage{color}
\usepackage{lineno,hyperref}
\usepackage{feynmp}
\usepackage{subcaption}
\usepackage[font=scriptsize]{caption}

\newcommand{\be}{\begin{equation}} 
\newcommand{\ee}{\end{equation}}  
\newcommand{\bea}{\begin{eqnarray}}  
\newcommand{\eea}{\end{eqnarray}}

\begin{document}

\title{Model Discrimination in Gravitational Wave spectra \\ from Dark Phase Transitions}

\author{Djuna Croon$^{a,c}$}, 
\affiliation{$^a$Department of Physics and Astronomy, Dartmouth College,
  Hanover, NH 03755, USA}

\author{Ver\'{o}nica Sanz$^b$,} 
\affiliation{$^b$Department of Physics and Astronomy, University of Sussex, Falmer Campus, Brighton, United Kingdom}
  
 \author{$\rm and$ Graham White$^c$} 
\affiliation{$^c$TRIUMF, 4004 Wesbrook Mall, Vancouver, B.C. V6T2A3, Canada}

\date{\today}

\abstract{ 
In anticipation of upcoming gravitational wave experiments, we provide a comprehensive overview of the spectra predicted by phase transitions triggered by states from a large variety of dark sector models. Such spectra are functions of the quantum numbers and (self-) couplings of the scalar that triggers the dark phase transition. We classify dark sectors that give rise to a first order phase transition and perform a numerical scan over the thermal parameter space. 
We then characterize scenarios in which a measurement of a new source of gravitational waves could allow us to discriminate between models with differing particle content.
}
\maketitle



\section{Introduction} 
The detection of gravitational waves (GW) \cite{Abbott:2016blz} established a new and independent probe of New Physics. 
It has already been suggested that the data from resolvable events such as from binary mergers could help constrain interacting dark matter \cite{Ellis:2017jgp,Croon:2017zcu} or exotic compact objects \cite{Giudice:2016zpa,Palenzuela:2017kcg,Croon:2018ftb}.
The observation also implies that we may anticipate the detection of a stochastic GW background at both current and future detectors. This may be a rare probe of the Cosmic Dark Ages and the first observational window onto cosmic phase transitions (PTs). Such cosmic phase transitions leave behind a characteristic broken power-law gravitational wave spectrum. The relic spectral shape depends on the strength of the transition, the speed of the transition, the bubble wall velocity and the temperature of the transition.

Most work thus far has considered the electroweak phase transition (EWPT), since a strongly first order electroweak phase transition can catalyze electroweak baryogenesis \cite{White:2016nbo,Morrissey:2012db,Trodden:1998ym,Dev:2016feu} providing an explanation for matter anti-matter asymmetry observed today \cite{deVries:2017ncy,Balazs:2016yvi,Akula:2017yfr,Balazs:2013cia,Balazs:2004ae,Lee:2004we,Morrissey:2012db,Trodden:1998ym}. The phenomenology of EWPTs have been studied abundantly \cite{Balazs:2016tbi,Jinno:2017ixd,Matsui:2017ggm,Huang:2017kzu,Baldes:2017ygu,Demidov:2017lzf,Chala:2018ari,Hashino:2018zsi,Grojean:2006bp,Caprini:2015zlo,Apreda:2001us} and the GW production has recently been advanced theoretically \cite{Hindmarsh:2016lnk} and through lattice simulations \cite{Hindmarsh:2017gnf,Cutting:2018tjt}. From these results, it has become clear that the successfulness of electroweak baryogenesis and the observability of the GWs from the EWPT are somewhat in tension. 

In terms of the LISA-inverse problem, much attention has been focused towards either arguing for a new scale of physics \cite{Grojean:2006bp,Dev:2016feu}, or for relic backgrounds from certain well motivated extensions of the standard model, assuming the reheating temperature is sufficiently high \cite{Balazs:2016tbi,Jinno:2017ixd,Matsui:2017ggm,Huang:2017kzu,Baldes:2017ygu,Demidov:2017lzf,Chala:2018ari,Hashino:2018zsi,deVries:2017ncy,Balazs:2016yvi,Akula:2017yfr,Balazs:2013cia,Balazs:2004ae,Lee:2004we,Morrissey:2012db,Trodden:1998ym,Croon:2018new,Schwaller:2015tja}. Little work to date has focused on the question of model discrimination \cite{Jinno:2017ixd}. In this work we endeavour to see how much model discrimination is in principle possible from the frequency spectrum of a future stochastic gravitational wave signal. In particular, we consider renormalizable and non-renomalizable effective field theories of interacting hidden sectors, in which a gauge symmetry is spontaneously broken. We also consider the effect of fermions that couple to the scalar. 

Simulations of gravitational wave backgrounds from cosmic phase transitions indicate that there are three spectral contributions: the collision spectrum is the direct effect of bubbles of true vacuum colliding, the sound wave spectrum is the result of the fluid dynamics after such collisions, and the turbulence spectrum, which is usually subdominant. It has been realized recently that the sound wave contribution dominates in most relevant scenarios. In particular, this is true in all cases that do not display "runaway" behaviour, and such runaway is blocked by any gauge bosons acquiring a mass in the transition \cite{Bodeker:2017cim}. 

All spectra are controlled by four thermal parameters: the velocity of the bubble wall, $v_w$, the ratio of the free energy density difference between the true and false vacuum and the total energy density, $\xi$, the speed of the phase transition $\beta/H$ and the nucleation temperature $T_N$. In the special case in which two peaks are visible, the four thermal parameters can in principle be reconstructed.  

In this paper we focus on the thermal parameters $T_N, \beta/H$ and $\xi$. The first determines the scale of the phase transition, and the latter two are most powerful at model discrimination. To study the thermal parameters in a general context, we observe that first order phase transitions are realized in (effective) double well potentials, from the interaction of terms with alternating signs. As such, we will study multiple models within two limiting scenarios, 
\bea \label{limiting234}    V(h_D,T) &=& \frac{1}{2}m(T)^2 h_D^2 - c_3(T) h_D^3 + \frac{1}{4} \lambda(T) h_D^4 \\ \label{limiting246}
    V(h_D,T) &=& \label{effV2} \frac{1}{2}m(T)^2 h_D^2  - \frac{1}{4} \lambda(T) h_D^4 +   c_6(T) h_D^6
\eea
where all coefficients are positive at the time of transition. Most phase transitions can be mapped onto these effective scenarios. In particular, 
the EWPT in a Higgs+singlet model is an example of \eqref{effV2}, upon integrating out the heavy singlet (up to dimension-6 operators).

The thermal parameters are strongly dependent on the nature of the thermal corrections. These thermal corrections are functions of the bosonic and fermionic degrees of freedom coupling to the scalar. The bosonic degrees of freedom are given by the gauge structure of the theory. 
Here we will consider models within the following scenarios, corresponding to the limiting cases \eqref{limiting234} and \eqref{limiting246}:
\begin{enumerate}
    \item A dark Higgs - $SU(N)$ breaking into $SU(N-1)$. In this case the barrier between the true and false vacuum during the transition is caused by dark gauge bosons that provide an effective cubic term.
    \item A dark Higgs - $SU(N)$ breaking into $SU(N-1)$ -  with significant non-renormalizable operators. In this case the barrier between the true and false vacuum is caused by the quartic dark Higgs coupling being negative and the vacuum being stabilized by the positive Wilson coefficient of the sextet interaction.
\end{enumerate}
Such scenarios may arise for example in the context of Composite Higgs models of cosmology \cite{Croon:2015naa,Croon:2015fza}, where the Dark Higgs would be represented by a pseudo-Goldstone boson state (a generalization of a QCD pion) whose interesting potential is due to explicit breaking of the global symmetry via $SU(N)$ gauge and Yukawa couplings.

In each case we consider gauge groups of different ranks as well as models with and without a thermal mass produced by dark fermions. For all cases $\xi$ is independent of the scale of the potential and $\beta/H$ has a weak logarithmic dependence whereas both thermal parameters are controlled by the ratio of the vev with the scale of the potential $x\equiv v/\Lambda$. Therefore the renormalizable potentials are a 2(3) parameter problems for each model and the non-renormalizable potentials are a 3(4) parameter problem without (with) the addition of a dark fermion. 

We find that non-renormalizable operators dramatically improve the visibility of gravitational wave spectra, whereas adding a dark fermions $N_f$ and increasing the rank of the group $N$ provide a more modest boost, which becomes reasonably large in the limit of large $N_f$ or $N$. The boosts to visibility in each case are non-degenerate. In the renormalizable case \eqref{limiting234}, we find that both the effect of a larger gauge group ($SU(N)\to SU(N+1)$), and the effect of increasing the number of fermions (with significant thermal mass) are essentially to shift the thermal parameter space, and increase the detection prospects. Of course, there is a degeneracy of predictions for specific models. 
It has been suggested that anisotropy measurements could break this degeneracy, for example by a cross-correlation with the CMB data  \cite{Geller:2018mwu}.

The structure of this paper is as follows. In section \ref{sec:models} we summarize the models we are attempting to discriminate. In section \ref{sec:GW} we review the spectra of gravitational waves from a cosmic phase transition and in section \ref{sec:results} we present our results. In section \ref{sec:DMabundance} we relate our results to studies of dark matter, before concluding with a discussion and an outlook to future work in the final section.

\section{Scenarios for a dark first order Phase Transition}\label{sec:models}

A first order phase transition may occur for a potential with three competing terms, with alternating signs, such that it has a double well separated by a barrier. 
Moreover, the vacuum energy corresponding to these minima will be temperature-dependent, such that the ground state changes as the Universe cools. 
The first order phase transition may then happen if the potential barrier is present at the critical temperature $T_c$, when the minima are degenerate.

We will consider two limiting cases of such potentials. In the renormalizable case, the potential barrier is generated effectively at finite temperatures, but does not exist at zero temperature. 
As we will see, the zero-temperature masses and self-couplings, the quantum numbers of the scalar, and the couplings to fermions crucially determine the thermal parameters of the phase transition. For all the models we are considering the part of the Lagrangian relevant to phase transitions can be written

\begin{eqnarray}
{\cal L} \supset | D_\mu H_D |^2- V(H_D)+\sum_i^{N_f} i \bar{\chi} _i D_\mu \gamma ^\mu \chi _i  -y_{\chi}^i v h_D \chi _i  \bar \chi _i 
\end{eqnarray}
where $\langle H_D \rangle = v$, $h_D$ is the singlet part of $H_D$, and the covariant derivative of the Higgs (but not necessarily of the fermion) contains the dark gauge bosons which are in the adjoint representation of SU(N), with gauge coupling constant $g$ and covariant derivative
\begin{equation}
    D_\mu = \partial _\mu - ig \tau ^a A^a _\mu  \ .
\end{equation} 
We will consider potentials of the form \eqref{limiting234} and \eqref{limiting246}.

\subsection{$SU(N)/SU(N-1)$ models with renormalizable operators}
The first case has a double well generated from the quadratic, cubic, and quartic interactions at finite temperature. We parametrize the potential such that the overal scale ($\Lambda$) and the zero temperature vacuum expectation value ($v$) are inputs. This implies the following redefinitions for zero temperature parameters in the potential \eqref{limiting234}, 
\bea m^2(0) &=& -  \frac{\Lambda^4}{v^2}, \\
\lambda(0) &=& \frac{\Lambda^4}{v^4}. \eea 
As we will see below, we find that some thermal parameters are only functions of the ratio of the zero temperature vev and the scale of the potential ($v/\Lambda$). 
 Using this parametrization, the finite temperature potential is given by,
\begin{eqnarray} \notag
V(H,T)&=&\Lambda^4\left[ -\frac{1}{2}\left( \frac{h_D}{v} \right)^2+\frac{1}{4} \left( \frac{h_D}{v} \right)^4 \right] \\ &&+
\frac{T^4}{2 \pi ^2} \left[\, \sum _{i \, \in\, \rm bosons} n_i J_B\left(\frac{m_i^2}{T^2}\right) +  \sum _{j \, \in\, \rm fermions} n_j J_F\left(\frac{m_j^2}{T^2}\right) \right] \nonumber \\
&=& \Lambda^4 \left[ \left(-\frac{1}{2} + \left(\frac{1}{8}+ \frac{N_G}{24} \right) \frac{T^2}{v^2} +\frac{3}{24}N_{GB} \frac{g^2}{4} \frac{T^2 v^2}{\Lambda^4}+y^2 N_f \frac{T^2}{24}\frac{v^2}{\Lambda^4} \right) \left( \frac{h_D}{v} \right)^2 \right. \nonumber \\ &&\left. - \left( N_{GB} \left( \frac{g^2}{4} \right)^{3/2} \frac{1}{4 \pi} \frac{v^3 T}{\Lambda ^4} \right)\left( \frac{h_D}{v} \right)^3+\frac{1}{4} \left( \frac{h_D}{v} \right)^4 \right], \label{234potential} 
\end{eqnarray}
where
$N_{G}$ is the number of gauge bosons coupling to the scalar sector with coupling constant $g$, $N_{GB}$ is the number of Goldstone degrees of freedom, and $N_f$ is the number of fermions with Yukawa coupling $y$. For simplicity, we consider degenerate Yukawa couplings, as the gravitational waves produced by $(y,N_F)$ and $(\{y_i\} ,N_F^\prime )$ are related by $y^2 \times N_F = \sum ^{N_F^\prime} y_i^2  $.\footnote{Up to a small change in the number of relativistic degrees of freedom $g^*$. Since the gravitational wave spectra has a very weak dependence on $g^*$, making this simplification is at little cost to generality.} In the second line we have applied a high temperature expansion, 
\bea \label{hightem} J_B \left(\frac{m^2}{T^2}\right) \sim \frac{m^2}{24T^2}-\frac{m^3}{12 \pi T^3}, \,\,\,\,\,\,\,\,\,\,\text{     and     }\,\,\,\,\,\,\,\,\,\, J_F\left(\frac{m^2}{T^2}\right) \sim -\frac{m^2}{48 T^2}.\eea All field dependent masses which enter into the effective potential are provided in the appendix.\footnote{Note that the use of perturbation theory introduces some theoretical uncertainty as perturbativity at finite temperature breaks down above the critical temperature \cite{Quiros:1999jp,Gorda:2018hvi}, a fact that can be delayed somewhat by the inclusion of "daisy terms" \cite{Curtin:2016urg} although in reality one requires a lattice simulation for a robust treatment. In spite of this theoretical uncertainty we expect our results to be indicative of the overall thermal parameter space including its overall scope and dependence on the model. Finally note that the most important points in our scan are where a lot of supercooling occurs and $T_C$ is significantly higher than $T_N$ meaning that these are the points where perturbation theory is most valid.}

\subsection{$SU(N)/SU(N-1)$ models with non-renormalizable operators}
\begin{figure}
    \centering
    \includegraphics[width=0.48\textwidth]{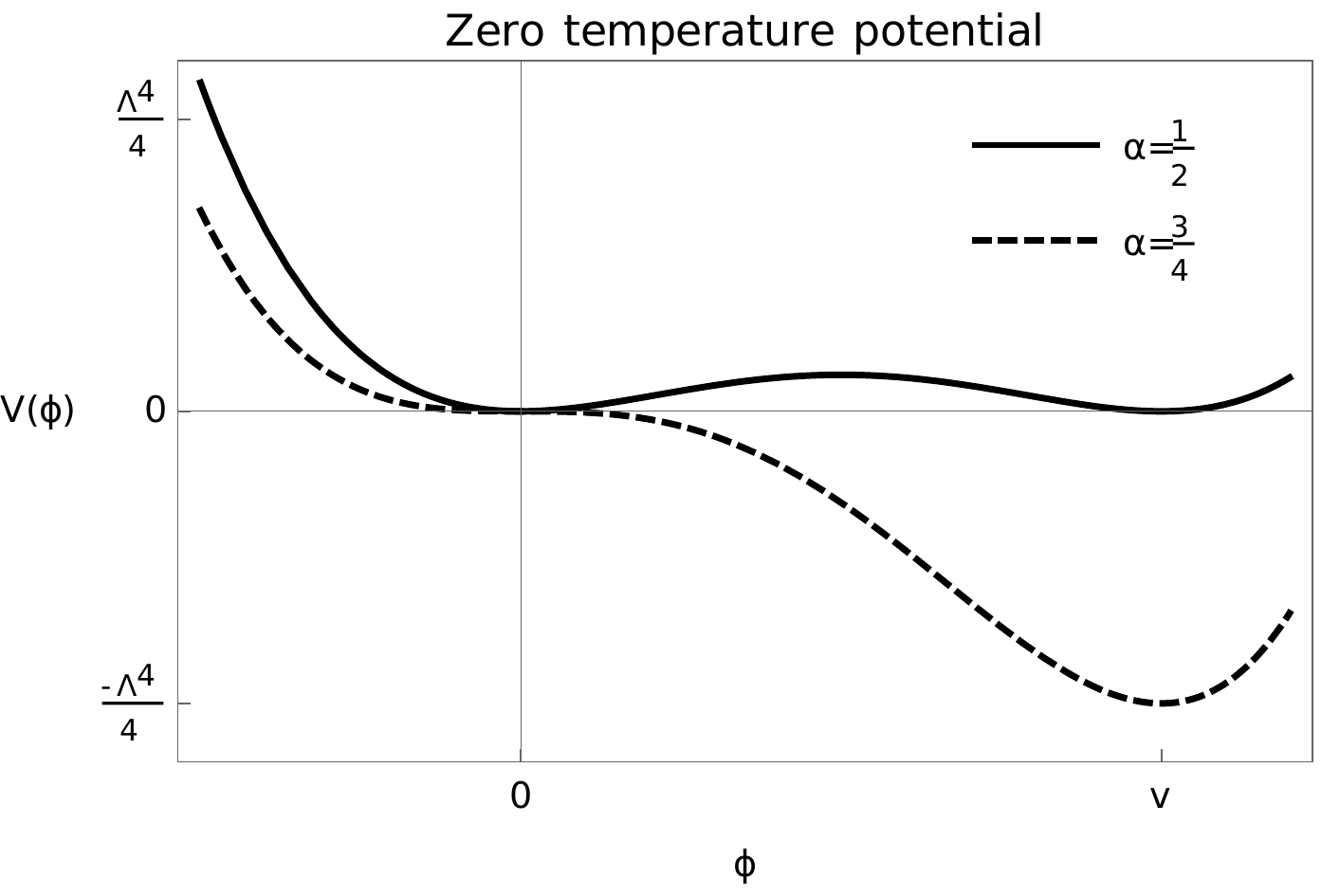}
    \caption{The potential \eqref{246potential} at $T=0$ for the two limiting values of the parameter $\alpha$.}
    \label{fig:potential}
\end{figure}

The second limiting case has the double well resulting from the interplay between the quadratic, quartic, and sextic terms. We again choose a parametrization of the potential such that the scale of the potential $\Lambda$ and the zero temperature vacuum expectation $v$ value are inputs. This will leave us with one free parameter $\alpha$, which parameterizes the difference in vacuum energy of the two minima at zero temperature. In the high temperature expansion \eqref{hightem}, the potential becomes,
\begin{eqnarray} \label{246potential}
V(H,T)&=& \Lambda ^4 \left[\left( 2-3\alpha -\left[\frac{1}{2} +N_G \frac{1}{6} \right]\frac{T^2}{v^2} +\frac{3}{24}N_{GB} \frac{g^2}{4} \frac{T^2 v^2}{\Lambda^4}+y^2 N_f \frac{T^2}{24}\frac{v^2}{\Lambda^4} \right) \left( \frac{h_D}{v} \right)^2 \right. \nonumber \\ &&\left. -\left(1-\frac{(30+6N_G) \alpha T^2}{v^2}\right) \left(\frac{h_D}{v} \right)^4 + \alpha \left( \frac{h_D}{v}\right)^6 \right] \ . \label{eq:V6T}
\end{eqnarray}
It is seen that at zero temperature, the potential has minima at $h_D=0$ and $h_D=v$ respectively, overall scale $\Lambda$, and the (dimensionless) non-renormalizable coupling is $\alpha$. That is, we have made the following redefinitions in Eq. \eqref{limiting246}:
\bea m^2(0) &=& (2 - 3 \alpha) \, \frac{\Lambda^4}{v^2} \\ 
\lambda(0) &=& 4 \, \frac{\Lambda^4}{v^4} \\ c_6 (0) &=& \alpha \, \frac{\Lambda^4}{v^6}\eea
At zero temperature, a value of $\alpha=1/2$ corresponds to degenerate minima, and the upper limit $\alpha=2/3$ corresponds to the value for which there is no zero temperature barrier between the vacua (as the zero temperature mass term changes sign), see Fig.~\ref{fig:potential}. Of course, finite temperature corrections may allow for a higher value of $\alpha$, as positive corrections to the mass term may reintroduce the barrier. Note we have once again assumed degenerate Yukawas with little loss of generality as explained in the previous section.

For operators up to dimension-6, models for the electroweak phase transition (EWPT) can be captured effectively by a special case of the above, with 
\begin{equation} \label{EWPTpotential}
  V_6 (h,T)= \left(a_T T^2 -\frac{\mu ^2}{2}\right) h^2 + \left(b_T T^2 -\frac{\lambda }{4}\right) h^4 + \frac{1}{8 \Lambda _ 6 ^2} h^6 
\end{equation}
where we have defined, 
\begin{eqnarray}
a_T&=&\frac{y_t^2}{8}+3\frac{g^2}{32}+\frac{g\prime {}^2}{32}-\frac{\lambda}{4} + \frac{v_0^2}{\Lambda _6^2}\frac{3}{4}  \\
b_T &=& \frac{1}{4} \frac{1}{\Lambda _ 6 ^2} .
\end{eqnarray}
Here $\Lambda_6$ is the scale associated with the dimension-6 operators which arise from integrating out BSM physics, such as a singlet scalar.\footnote{The notation in \eqref{EWPTpotential} may differ from the literature, in that we use $\lambda>1$ such that the zero temperature form of the potential is explicit.}. We will also consider the EWPT with non-renormalizable operators for the sake of comparison later. \par
Finally, note that if one rewrites Eq. \ref{eq:V6T} in terms of implicitly defined temperature dependent parameters
\begin{equation}
    V(H,T)= \Lambda (T)^4 \left[ \left(2-3 \alpha (T) \right) \left( \frac{h}{v(T)} \right)^2 -\left( \frac{h}{v(T)} \right)^4 + \alpha (T) \left( \frac{h}{v(T)} \right)^6 \right]
\end{equation}
one can follow the process in \cite{Akula:2016gpl,Croon:2018new} and fit the action to the function for the range $\alpha (T) \in [0.51,0.65]$
\begin{equation}
    S_E = \frac{v(T)^3}{\Lambda (T)^2} 10^{\sum _{i=1} ^3 a_i (\alpha (T) - 2/3)^i}
\end{equation}
with $a_i = (-17.446,-132.404,-763.744)$. 

\section{Mapping onto Dark Matter models}\label{sec:Mapping}

In this section we give examples of hidden sector models which can be mapped onto our general framework given above. Of course we are not completely general as we do not consider for example the case where multiple scalars acquire a vev at the same time (such as a multi dark higgs doublet model) or more complicated gauge group structures SU(N)$\times$SU(N$^\prime$) where both gauge couplings are large. However, scenarios which can be mapped onto our framework are ubiquitous including, Pati-Salem symmetry breaking\footnote{This phase transition is more likely to occur at a scale visible to aLIGO than LISA} \cite{King:2014iia}, colour breaking intermediate phase transitions \cite{Ramsey-Musolf:2017tgh,Patel:2013zla}, atomic dark matter \cite{Kaplan:2009de}, asymmetric dark matter \cite{Petraki:2011mv} and compositeness \cite{Ma:2017vzm} to give a non-exhaustive list. We give more details of three of these examples and how they map to the various models we consider below.

\subsection{Generalized baryon number}
As was suggested in \cite{Petraki:2011mv}, the dark sector relic abundance and the baryon asymmetry in the SM can have a common origin in models with a generative symmetry breaking. In such models, there is a generative gauge group G, for example $SU(2)_G$ which is broken spontaneously through a first-order phase transition in the early universe. The asymmetry generated in this phase transition is communicated to the dark and visible sectors through a mixed Yukawa term.
The degenerative scalar has tree-level zero temperature potential, 
\bea V(\varphi) = -\frac{\mu^2_\varphi}{2} |\varphi|^2 + \lambda_\varphi |\varphi|^4 \eea
and quartic mixing terms with the SM Higgs, B-L breaking scalar $\sigma$, and dark scalar $\chi$. For small mixing, such as is the case in various supersymmetric models, the mass contributions are small. For non-supersymmetric models, the mixing can be significant, and contribute to the thermalization and decay properties of the various sectors. The mass hierarchies are small, such that the scalar $\varphi$ can have a mass at the electroweak scale. In this case there are significant cosmological and astrophysical constraints as discussed in \cite{Petraki:2011mv}. 

The first order PT can be induced when one includes an effective dimension-6 operator, which can arise at the one loop level from the mixed quartic interactions \cite{Baldes:2017ygu}
\bea V_6(\varphi) = \frac{1}{8 \Lambda^2} \phi^6 \eea
from which it is seen that this an example of a model within the scenario given by \eqref{246potential}.
\subsection{Atomic dark matter}
A further possibility is that the dark sector contains a confining group, as well as fermions charged under an unbroken $U(1)'$. Then, dark atoms can be formed \cite{Kaplan:2009de}. The strongest constraint on atomic dark matter comes from the self scattering bound \cite{Ackerman:mha,Agrawal:2017rvu},
\begin{equation}
\frac{\alpha _D^2}{m_\chi ^3} \leq 10^{-11} ({\rm GeV} ^{-3}) \ 
\end{equation} 
where $m_ \chi $ is the heavier particle, which forms the nucleus of dark atoms. The mass of $m_\chi$ can be heavier than a TeV \cite{Buckley:2017ttd} in which case the constraint on the gauge coupling is very modest ($\alpha _D\sim 0.1$, implying $g \sim \mathcal{O}(1)$). A simple example is an SU(4) gauge group, which breaks into SU(3) $\times $U(1), allowing for the formation of nuclei during dark BBN \cite{Krnjaic:2014xza}. 

\subsection{Composite Dark Matter models}
A final example is a dark matter candidate as the lightest bound state of a confining gauge group $SU(N)$, such as has been discussed in \cite{Schwaller:2015tja}. The spontaneous symmetry breaking of an approximate global symmetry, which is only partially gauged, gives rise to pseudo-Goldstone bosons. These light states are sensitive to an effective scalar potential at the 1-loop level, which in turn initiates a further breaking. 
A particularly interesting possilibility has the SM Higgs and the dark matter candidate both as pseudo-Goldstone bosons of the same symmetry breaking \cite{Ma:2017vzm}.  Various symmetry breaking cosets have been studied in the literature, with scalar potentials of the form \eqref{234potential} or \eqref{246potential}. The couplings in such scenarios correspond to 1-loop integrals in the UV theory. The GW spectra for benchmarks of thermal parameters for the breaking $SU(3)$ and $SU(4)$ dark gauge symmetries were previously considered in \cite{Schwaller:2015tja}, where it was argued that scalar DM bound states and dark quarks (carrying EW quantum numbers) are most relevant for detection at LISA.

\section{Gravitational Waves from Phase Transitions}\label{sec:GW}
\subsection{Thermal parameters}
The dynamics of the phase transition are controlled by a bounce solution $\phi_{\rm c}( r,T)$, which is a spherically symmetric classical solution to the Euclidean equations of motion \cite{Coleman:1977py,Wainwright:2011kj,Akula:2016gpl}
\be\label{eomsinglet} - \frac{2}{r} \frac{\partial h_D }{\partial r}  - \frac{\partial ^2 h_D}{\partial r^2} + V'(\phi ) =0  \ . \ee
We compute the bounce solutions with potentials in the previous section. The thermal parameters of the phase transition can then be computed from the bounce solution. 

First, the nucleation temperature of bubbles of the new vacuum $T_N$ is conventionally defined as the temperature for which a volume fraction $e^{-1}$ is in the true vacuum state. This corresponds approximately to
\be p(t_N) t_N^4 = 1 \ee
where $p(t)$ is the nucleation probability per unit time per unit volume, and where $t_N$ is the nucleation time. The nucleation probability can be calculated from the bounce solution as,
\be p(T) = T^4 \,e^{-S_E/T} \ee
where $S_E$ is the Euclidean action evaluated on the bounce. We assume a radiation dominated universe to relate the nucleation temperature and time. 
The speed of the phase transition is controlled by the parameter $\beta$, which can also be related to the bounce action,
\begin{equation}
    \frac{\beta}{H} \sim\left. T \, \frac{d (S_E/T)}{dT} \right|_{T=T_N} \ .
\end{equation}
Last, the latent heat parameter is given by,
\begin{equation}
    \xi = \left.\frac{1}{\rho_N}\left( \Delta V - T \Delta \frac{dV}{dT}\right)\right|_{T=T_N} \ .
\end{equation}
Where $\Delta $ indicates that the quantity should be evaluated on both sides of the bubble wall, and where $\rho _N= \pi ^2 g^* T_N ^4/30$ is the equilibrium energy density at $T_N$.

\subsection{Gravitational wave spectrum and the LISA inverse problem}
The gravitational wave profiles can be related to the thermal parameters. We will adopt a parametrization introduced by \cite{Weir:2017wfa}, but our analysis can be adapted when future models become available. 
In principle, there are three contributions to the power spectrum, \be\label{OmegaGWoneT} \Omega_{\rm GW} = \Omega_{\rm col} + \Omega_{\rm sw} + \Omega_{\rm turb} \ee
Where the first term corresponds to the spectrum from bubble collisions, the second is a spectrum due to sound waves in the fluid after collisions, and the third a turbulence term. 

\begin{figure}
    \centering
    \includegraphics[width=0.68\textwidth]{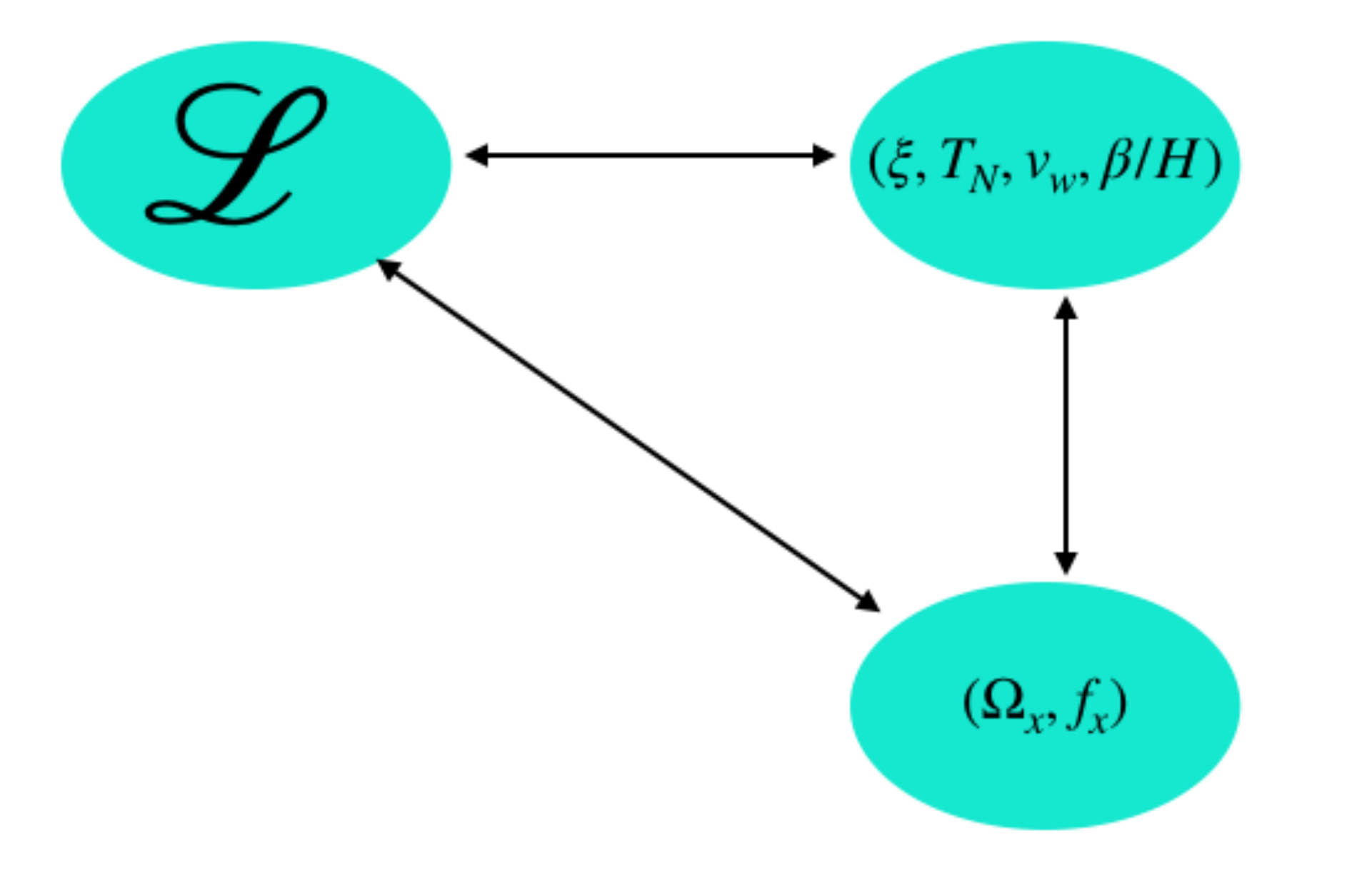}
    \caption{\emph{(Schematically) the LISA inverse problem}. In the above, the subscript $x$ refers to the dominant peak of the GW spectra (collision, sound wave, or turbulence). As described in the text, for most models the sound wave contribution is dominant. The thermal parameters of the PT can be calculated by solving the bounce EOM \eqref{eomsinglet}, and then related to the GW spectra using \eqref{ampsw} and \eqref{freqsw}. This paper finds general relations between the GW spectra and the Lagrangian.}
    \label{fig:lisainverse}
\end{figure}

As realized last year \cite{Bodeker:2017cim}, in any model in which gauge bosons gain a mass in the transition, the bubble wall velocity approaches a finite limit. Therefore, the sound wave contribution \cite{Hindmarsh:2016lnk} is typically dominant in all of the cases we consider in this work. Its power spectrum can be expressed as \cite{Weir:2017wfa},
\begin{equation}\label{ampsw}
h^2\Omega _{\rm sw}    = 8.5 \times 10^{-6} \left( \frac{100}{g_*} \right)^{-1/3} \Gamma ^2 \bar{U}_f^4  \left( \frac{\beta}{H} \right)^{-1}  v_w S_{\rm sw}(f) 
\end{equation}
where $\Gamma \sim 4/3$ is the adiabatic index, and $\bar{U}_f^2\sim (3/4) \kappa _f \, \xi$ is the rms fluid velocity.  For $v_w \rightarrow 1$, the efficiency parameter is well approximated by \cite{Espinosa:2010hh}
\begin{equation}
    \kappa _f \sim \frac{\xi }{0.73+0.083 \sqrt{\xi } +\xi}
\end{equation}
For $v_w \approx 0.5 $, we use \cite{Espinosa:2010hh}
\begin{equation}
    \kappa _f \sim \frac{\xi^{2/5} }{0.017+(0.997 + \xi)^{2/5}}
\end{equation}
and the spectral shape is given by
\begin{equation} 
    S_{\rm sw} =  \left( \frac{f}{f_{\rm sw}} \right) ^3 \left( \frac{7}{4+3\left( \frac{f}{f_{\rm sw}}\right) ^2} \right)^{7/2}
\end{equation}
with
\begin{equation} \label{freqsw}
    f_{\rm sw} = 8.9 \times 10^{-8} \, {\rm Hz}\, \left(\frac{1}{v_w}\right) \left( \frac{\beta}{H} \right) \left( \frac{T_N}{{\rm GeV}} \right) \left( \frac{g_* }{100} \right)^{1/6} \ .
\end{equation}
From this we notice that the amplitude of the signal is a function of the parameters $\beta / H$, the wall velocity $v_w$, and the latent heat $\xi$; whereas the position of the peak depends on $\beta / H$ and $T_N$. We will use this insight in the next section, to compare the predictions of the different models \eqref{234potential} and \eqref{246potential}. This effort can be summarized by the LISA inverse problem, in fig. \ref{fig:lisainverse}. 

We should mention some previous work towards solving the LISA inverse problem. The link between gravitational waves detection of collision and turbulence peaks and the thermal parameters has previously been summarized in ref. \cite{Grojean:2006bp}, which highlighted visible regions in the thermal parameter space. On the link between the Lagrangian and thermal parameters some thorough work has been done in the case of the EWPT with extended scalar sectors, \cite{Huber:2007vva,Delaunay:2007wb,Chala:2018ari,Vaskonen:2016yiu}. The aim of this paper is to compliment these previous works by studying the general case of a (single) scalar, with couplings to different numbers of fermions and gauge bosons, as well as other scalars separated in mass. 

\section{Spectra from models}\label{sec:results}
We compute the thermal parameters for scenarios \eqref{234potential} and \eqref{246potential}, for different dark sectors. We are specifically interested in light scalar sectors, with masses around the EW scale. For comparison, we also study the SMEFT case, in which the electroweak phase transition is catalyzed by a non-renormalizable $H^6$ effective operator. The SMEFT case is then well approximated by a dark SU(2) with three dark fermions. 

We find bounce solutions using two techniques to ensure accuracy: a numerical finite-difference algorithm, where we discretized the radial direction $r$ and the analytic technique described in section \ref{sec:models}. The thermal parameters are then found by substituting the bounce solution into the Euclidean action $S_E$ as described in the previous paragraph. 

In both the renormalizable and non-renormalizable models the thermal parameter set ($\xi, H/\beta $) governs the peak amplitude. We find that these results are essentially independent of the scale of the potential $\Lambda$. Specifically, $\xi$ is independent, whereas $\beta/H$ has a weak Logrthmic dependence. The nucleation temperature by contrast scales linearly with $\Lambda$. In the case where we have only renormalizable operators \eqref{234potential}, we scan over $(g,v/\Lambda)$, with scan ranges $g \in (0.1,1)$, and $v/\Lambda \in (0.5,4)$
In the non-renormalizable case \eqref{246potential}, we fix $g=(0.5,1)$ and scan over $(\alpha, v/\Lambda )$, where we fix $\Lambda =200$ GeV. The scan ranges are $v/\Lambda \in (0.5,4)$ and $\alpha \in (0.55,1.5)$. 
We assume that the fermions are massless before the PT. The parameter that enters the scan is then $N_f \times y_\chi$. For convenience, we have set $y_\chi =1$ in the figures. 

\begin{figure}
    \centering
    \includegraphics[width=0.48\textwidth]{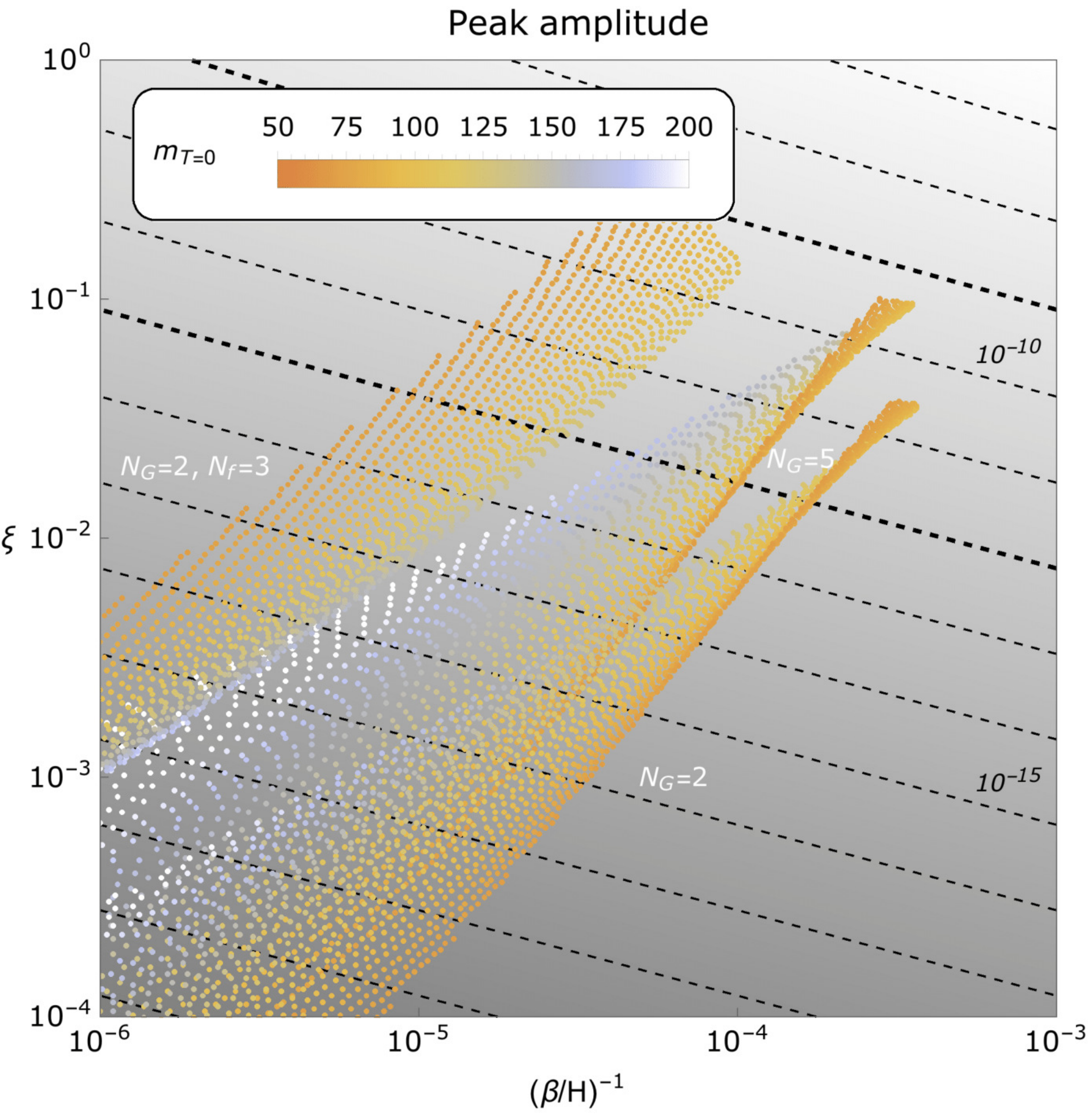}
     \includegraphics[width=0.48\textwidth]{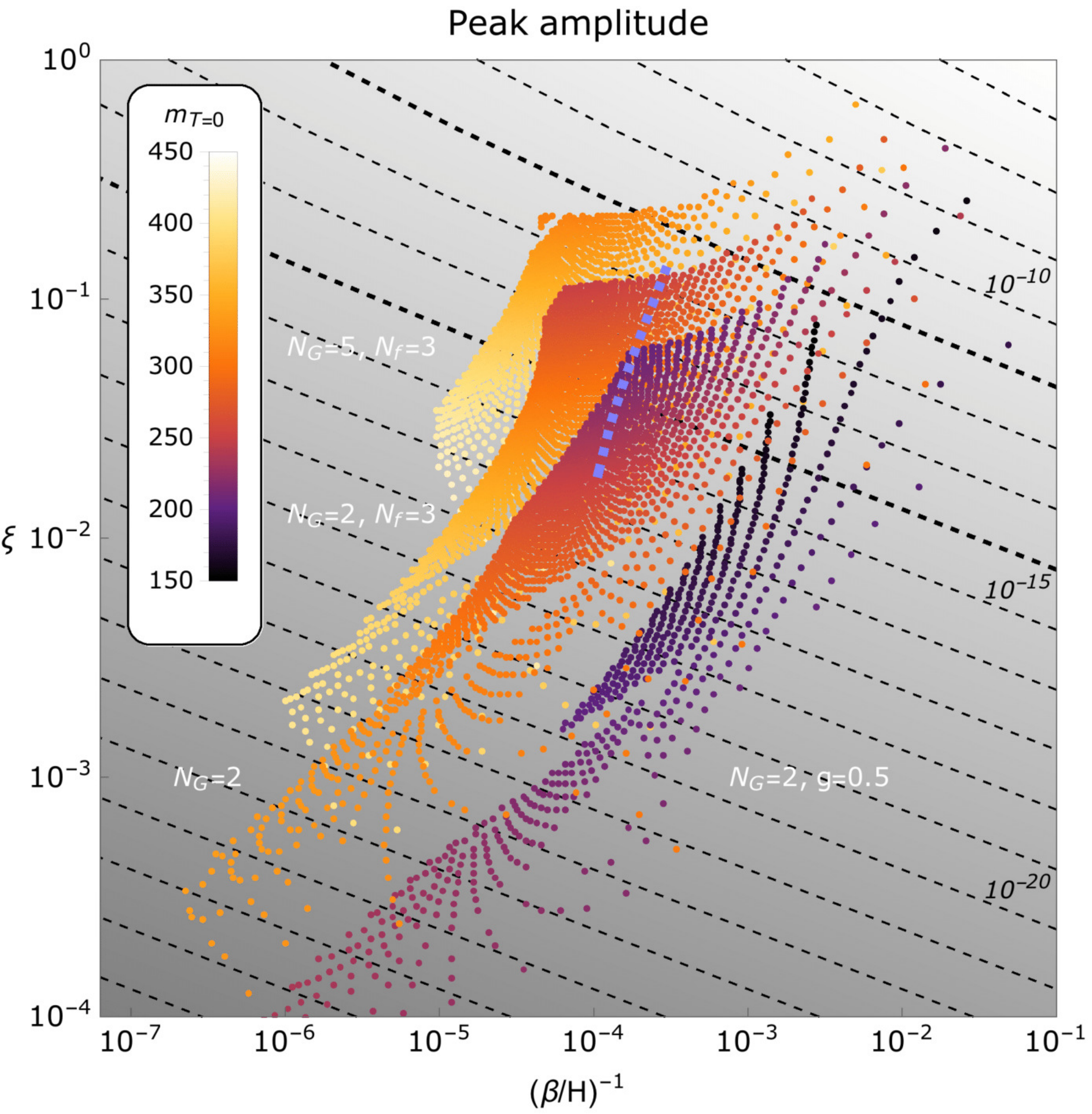}
    \caption{Thermal parameters from the PT described by Eqs \eqref{234potential} (left) and \eqref{246potential} (right). The dominant contribution to the spectrum comes from the sound-wave term. The plot points are coloured by their effective zero temperature mass, found from $m^2 =d^2V/dT^2$ evaluated at $v$.
    The dashed contours in the plots correspond to the GW amplitude $\Omega_{\rm sw}$ \eqref{ampsw}, where we have chosen $v_w = 0.5$ in the left plot, and $v_w=1$ in the right plot (with the corresponding efficiencies from \cite{Espinosa:2010hh}), as motivated using the conditions in \cite{Bodeker:2009qy}. The upper thicker contour corresponds to the LISA 1-year peak sensitivity \cite{Moore:2014lga}. The lower thicker dashed contour corresponds to LISA for a power-law spectrum (integrated over frequency), taken from \cite{Thrane:2013oya}.
    The width of the contours is found from varying the zero-temperature potential parameters. Left: unless otherwise indicated, the number of Yukawa couplings is taken to be zero. If present, the Yukawa couplings are set to $y_\chi=1$. Right: unless otherwise indicated, $g=1$. The light blue dashed line corresponds to the predictions from the EWPT.}
    \label{fig:amps}
\end{figure}

\begin{figure}
    \centering
    \includegraphics[width=0.48\textwidth]{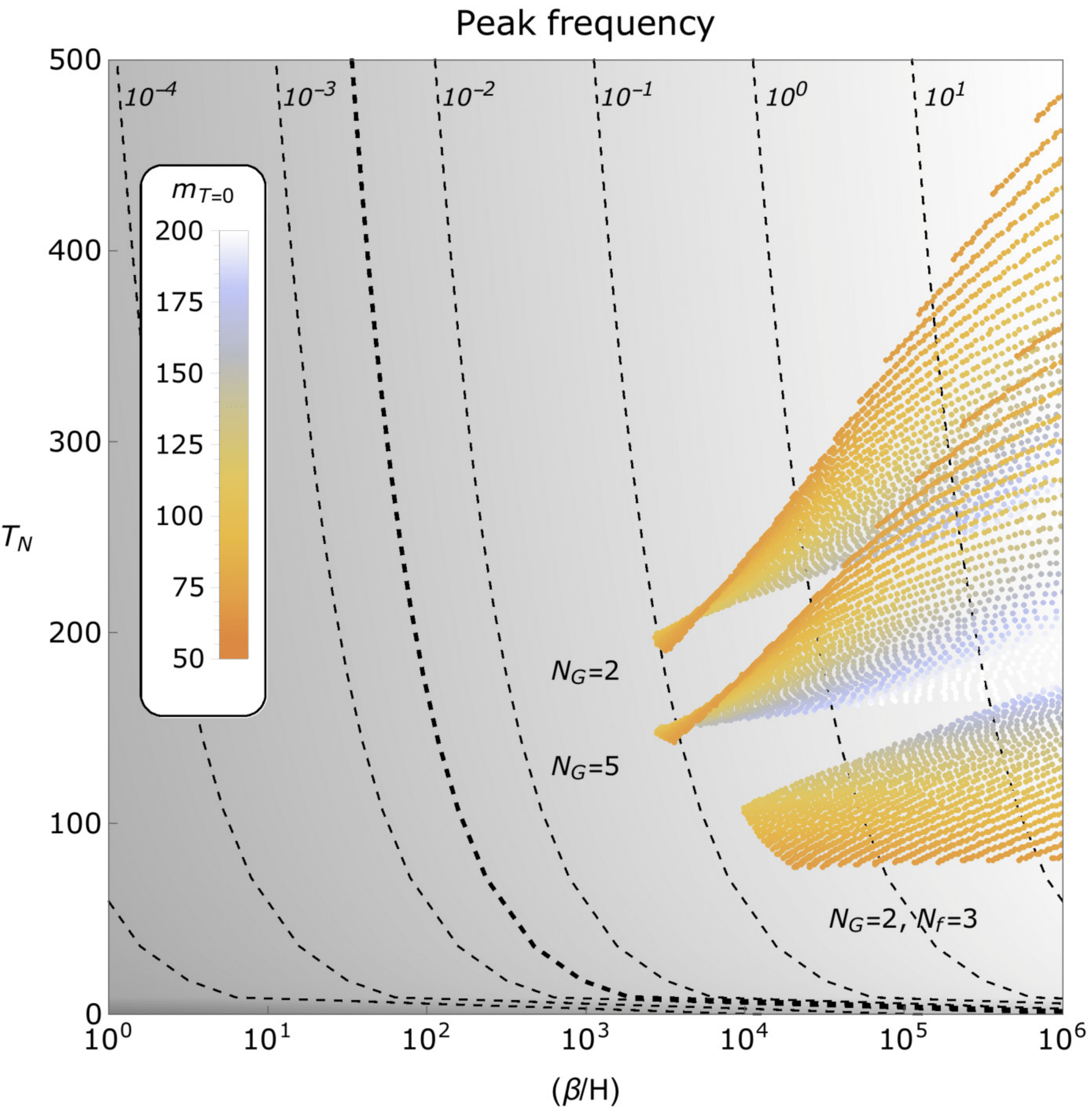}
    \includegraphics[width=0.48\textwidth]{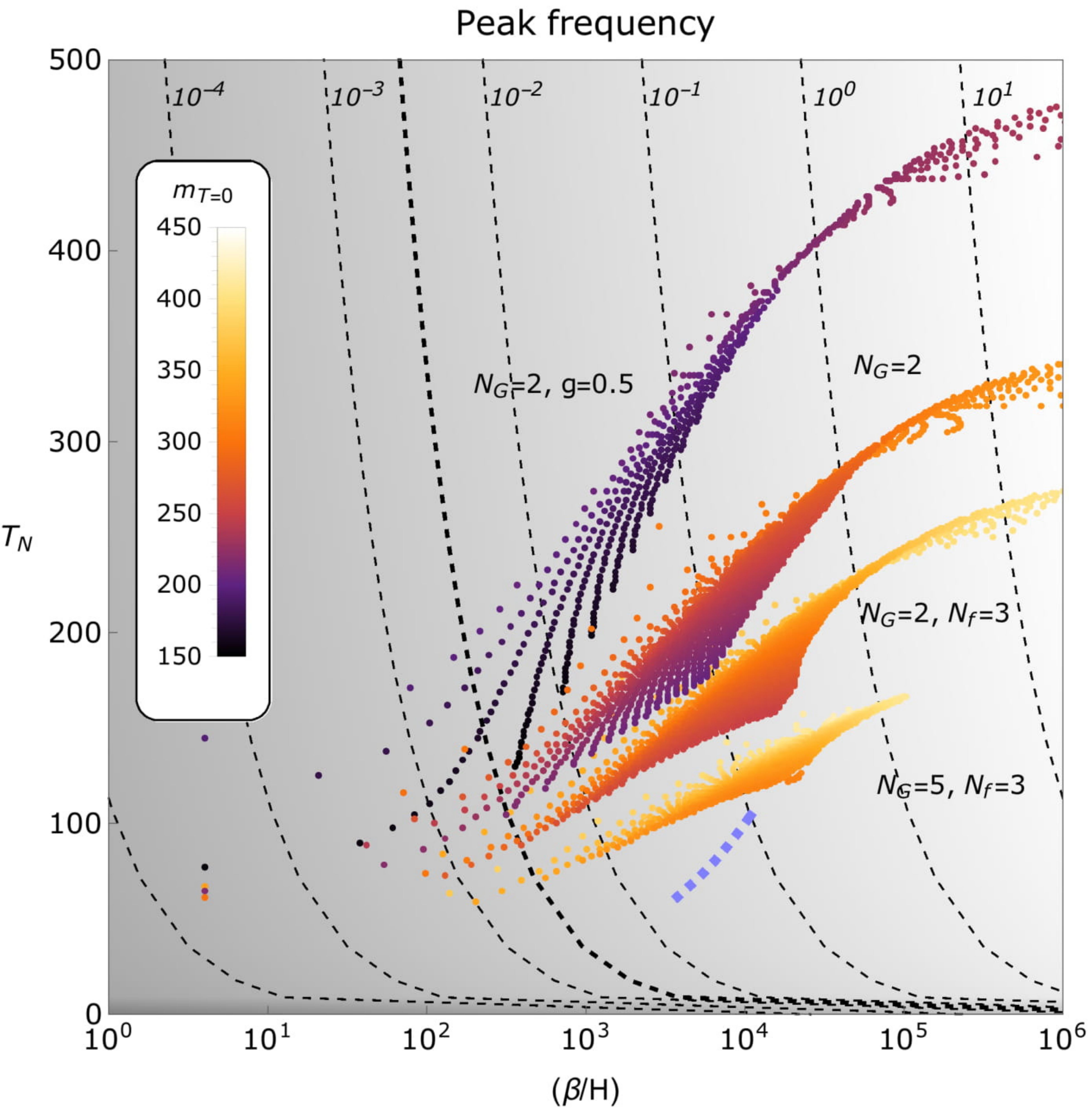}
    \caption{Thermal parameters from the PT described by Eqs \eqref{234potential} and \eqref{246potential} respectively. The dashed contours in the plots correspond to the sound wave peak $f_{\rm sw}$ \eqref{freqsw}, where we have chosen the wall velocities as in Fig. \ref{fig:amps}. The thicker dashed contour corresponds to the LISA frequency peak \cite{Thrane:2013oya}.  
    Note that the EWPT results do not overlap with our scans, since the nucleation temperature $T_N$ is sensitive to the scale $\Lambda$.}
    \label{fig:freqs}
\end{figure}

We summarize the results for the peak amplitude and peak frequency in Figs. \ref{fig:amps} and \ref{fig:freqs} respectively where in the spirit of reference \cite{Grojean:2006bp} we include visibility curves for LISA and plot the $(\xi, \beta/H)$ and $(\beta/H,T_n)$ planes.
We check explicitly that the high temperature expansion is valid for the results of our scan, by ensuring that $2 m_i^2<T^2$ with $i=h, \,GB$ for the gauge boson and Higgs mass at the critical vev and temperature.\footnote{We do not perform a similar check of the dark fermion mass, as (up to a change in the number of degrees of freedom $g^*$) the effect of dark fermions is a thermal mass controlled by $N_f\times y_\chi ^2$. That is, a reduction of the coupling to comply with the high temperature expansion can be compensated by increasing $N_f$.}
The effect of excluding points for which this check fails is to mildly trim the very tip of the peaks of the thermal parameter space in Figs. \ref{fig:amps}. The fact that the trimming occurs for low dark Higgs mass can be understood in direct analogy with early studies of the EWPT (before the Higgs mass was known). In this model one finds that for fixed vev, the strength of the phase transition grows inversely with the Higgs mass. In the limit of small Higgs mass, the gauge boson masses (which scale with $v(T_n)$) become large, invalidating the high temperature expansion ($m_G/T<1$) to be valid. 

The different shape of the results for the potential \eqref{234potential} with fermions can be understood as the fermions contribute only to the mass term. Therefore the potential barrier is no longer just a function of the gauge coupling, which we scan over, and the zero temperature mass. The reader will also notice that the results for the different potentials \eqref{234potential} and \eqref{246potential} have different zero temperature mass ranges. This can be understood by considering the contribution of the dimension-6 term to the latter. 

From the results for the non-renormalizable operators, it would naively seem that gauge bosons and fermions change the zero temperature mass of the scalar. However, the more accurate statement is that the presence of fermions and the rank of the gauge group determines which zero temperature masses lead to a strong first order PT, and are not disallowed by supercooling. Furthermore, for the case where $g=0.5$ rather than $g=1$, the high temperature expansion is valid for lower dark Higgs masses, before it is rendered invalid by large gauge boson masses.

In the right panels, we compare our result to the predictions from the EWPT up to dimension-6 operators \eqref{EWPTpotential}, with the dashed blue line. We find that the results in Fig. \ref{fig:amps} overlap, demonstrating that these results are insensitive to the scale $\Lambda$ (but sensitive to the ratio $v/\Lambda$). As expected, the predictions for the peak frequency (Fig. \ref{fig:freqs}) do not overlap, as $T_N$ scales with $\Lambda$.

\begin{figure}
    \centering
    \includegraphics[width=0.4\textwidth]{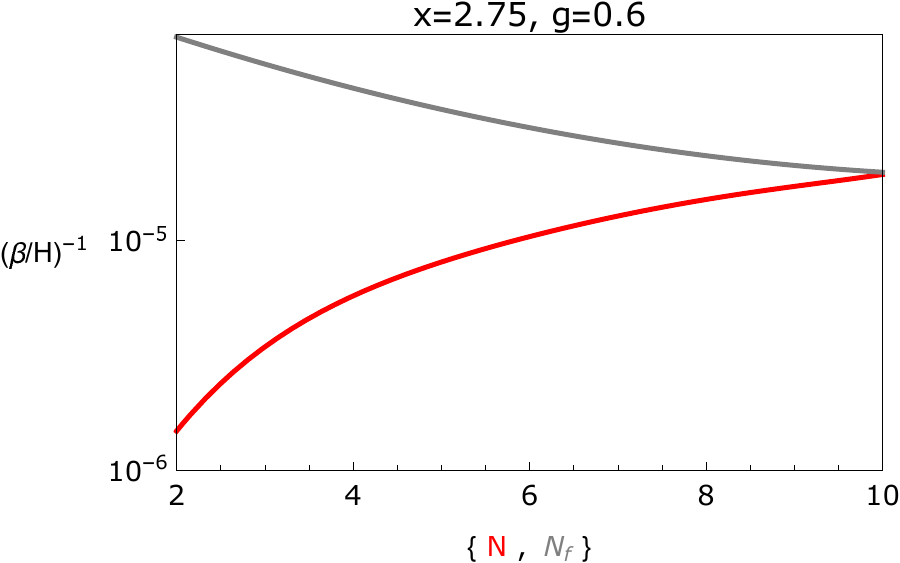}
     \includegraphics[width=0.4\textwidth]{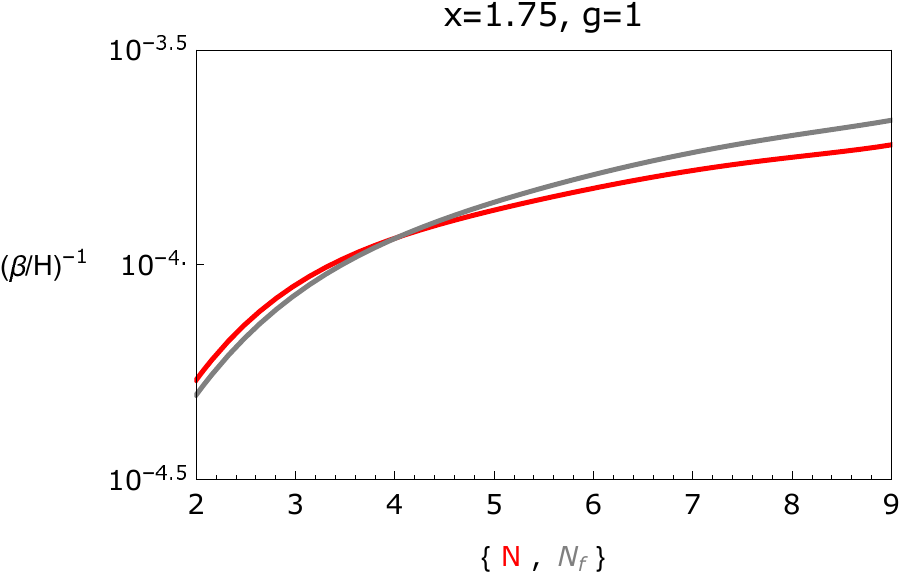}\\
     \includegraphics[width=0.4\textwidth]{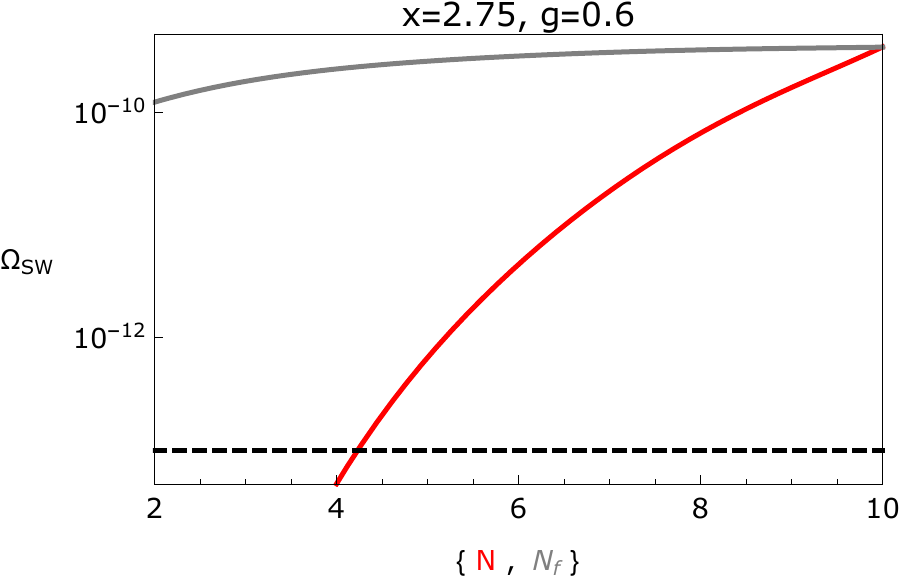}
     \includegraphics[width=0.4\textwidth]{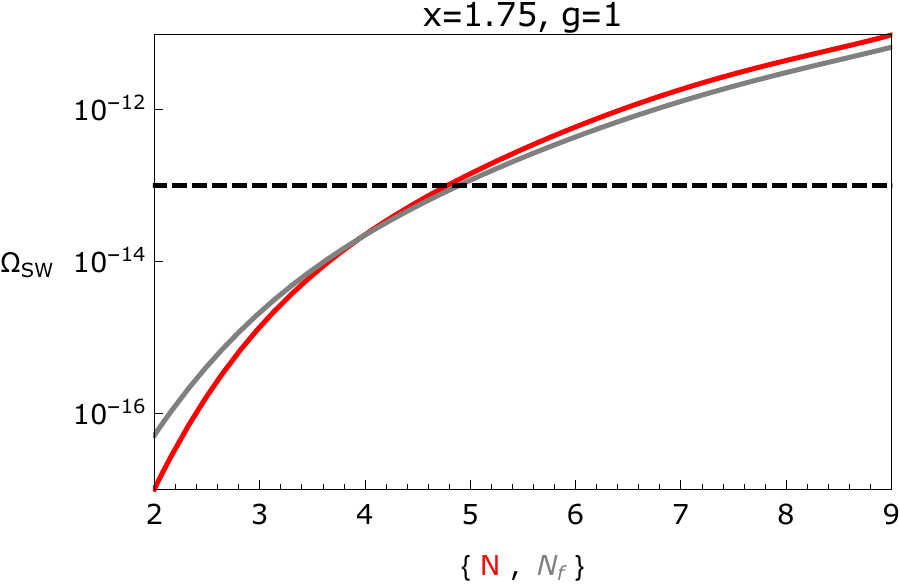}
    \caption{Thermal parameters from the PT described by Eqs \eqref{234potential} (left) and \eqref{246potential} (right). For the non-renormalizable potentials we take $\alpha =0.8$. In the cases where we vary $N_F$ we set $N=10(4)$, whereas in the cases where we vary $N$ we set $N_F=10(4)$ for the renormalizable and non-renormalizable potentials respectively. The black dashed line shows LISA visibility for a power spectrum integrated over frequency taken from \cite{Thrane:2013oya}. }
    \label{fig:correlations}
\end{figure}

Some qualitative features can lead to model discrimination, which we list below:
\begin{enumerate}
   \item  The thermal parameter space available for $SU(N)$ is essentially the same as that of $SU(2)$ apart from a shift in $\log \xi$ by an amount 
    \begin{equation}
        \Delta \log \xi  \sim  A(y_\chi \times N_f)\sqrt{N-2}
    \end{equation}
    where the coefficient $A(y_\chi \times N_f)$ depends on $y_\chi \times N_f$ and is around $2.4$ for $y_\chi\times N_F \sim 0$ and decreases to about 1.8 for $y\times N_f =10$. Note that in general increasing the rank of the gauge group improves visibility although one has diminishing returns for large $N$ which we show in Fig. \ref{fig:amps}.
    \item Adding fermions qualitatively changes the available thermal parameter space slightly. Comparing $N_f\times y _\chi  > 0$ and $N_F\times y_\chi=0$, we notice a shift and a slight change in shape. For $1<y\times N_f < 10$ we find that the thermal parameter space merely shifts according to
    \begin{equation}
        \Delta \log \xi = B(N) \sqrt{y_\chi \times N_f -1}, \quad         \Delta \log H/\beta = C(N) \sqrt{y_\chi \times N_f -1}
    \end{equation}
    where we find that $C(2)\sim-0.35$ for SU(2) and $C(10) \sim -0.3$ for SU(10)
    \begin{eqnarray}
    B(2) \sim 0.40, \quad B(5) = 0.23, \quad B(10) = 0.18
    \end{eqnarray}
    That is $\xi$ is shifted in the direction of greater visibility whereas $\beta/H$ is shifted in a direction of weaker visibility. Since the amplitude is more sensitive to $\xi$ this overall means that adding fermions increases the visibility of the transition which we show in both Fig. \ref{fig:correlations} and Fig. \ref{fig:amps}. The increase in $\beta/H$ is due to $T_c-T_n$ reducing in magnitude as one adds strongly coupled fermions. For $\xi$ there is a competition between two effects: the reduction in $T_c-T_n$ which tends to reduce $\xi$ and an increase in $dV/dT$ which increases $\xi$. Its the latter that wins.
    \item The presence of nonrenormalizable operators boosts $H/\beta$ by orders of magnitude compared to what is possible in the renormalizable case. This is a striking signal suggesting that a large $H/\beta$ indicates the presence of more than one new scale of physics. In this case the effect of adding extra fermions is to shift and slightly rotate  the thermal paramater space ($\xi, H/\beta$), this time in the $\Delta \log \xi$ direction although the relationship is less clean than the case of renormalizable operators. In contrast the effect of increasing the rank of the group is to both shift and somewhat contract the parameter space. The shift in both cases is in a direction of increased visibility. 
\end{enumerate}

\section{Relic abundance example} \label{sec:DMabundance}
The scenarios discussed in the previous sections constitute hidden sectors, which may explain the present relic abundance of Dark Matter (DM).
As an example, we discuss the contribution to DM relic abundance from the coupling to a single Dirac fermion to the scalar responsible for the PT. We will also assume the region $m_{h_D} < m_\chi$, which corresponds to the majority of the scenarios we covered in the last section. 

The fermionic DM may not have tree-level couplings to the SM, just Yukawa interactions with the Dark Higgs, and thermalize at a dark temperature, which in principle could be different from the SM evolution, $T_D \neq T_{SM}$. 
But provided that there was thermal equilibrium between the SM and hidden sector at some scale (above the weak scale), one can assume that at freeze-out of the $\chi$ particles $T_D \sim T_{SM}$. This scenario can explain the observed DM relic abundance \cite{Chacko:2015noa}, which is 
  mostly determined by the internal dynamics of the hidden sector. In particular, the annihilation $\bar\chi \chi \rightarrow h_D h_D$ sets the relic abundance of $\chi$ particles.

To avoid over-closure, the $h_D$ scalar is expected to have a decay channel to the SM, such as via Yukawa couplings to the SM fermions, via a mixing $\theta$ with the Higgs, of magnitude $g_f =  (m_f/v) \sin\theta$, where $y_\chi^2 \sin^2\theta \gtrsim 2 \times 10^{-13}$ \cite{Krnjaic:2015mbs}. This coupling is small enough such that the SM fermions are not expected to play a significant role in the $h_D$ phase transition.  

Under these assumptions, the dominant annihilation cross section is p-wave, and an approximate expression for the relic abundance is then given by~\cite{Lee:2013bua}
\begin{eqnarray}
\Omega_{DM} h^2 \simeq \frac{2.1\times 10^8 \textrm{ GeV}^{-1}}{M_P \sqrt{g_{\star} (x_F)} 3 b/x_F^2} \simeq 0.1 ,
\end{eqnarray}
where the fermion masses are $m_\chi = y_\chi v/\sqrt{2}$, $x_F= m_\chi/T_F \simeq 20$ and $b=(3/128 \pi) y_\chi^4/m_\chi^2$.

\begin{figure}
    \centering
    \includegraphics[width=.6\textwidth]{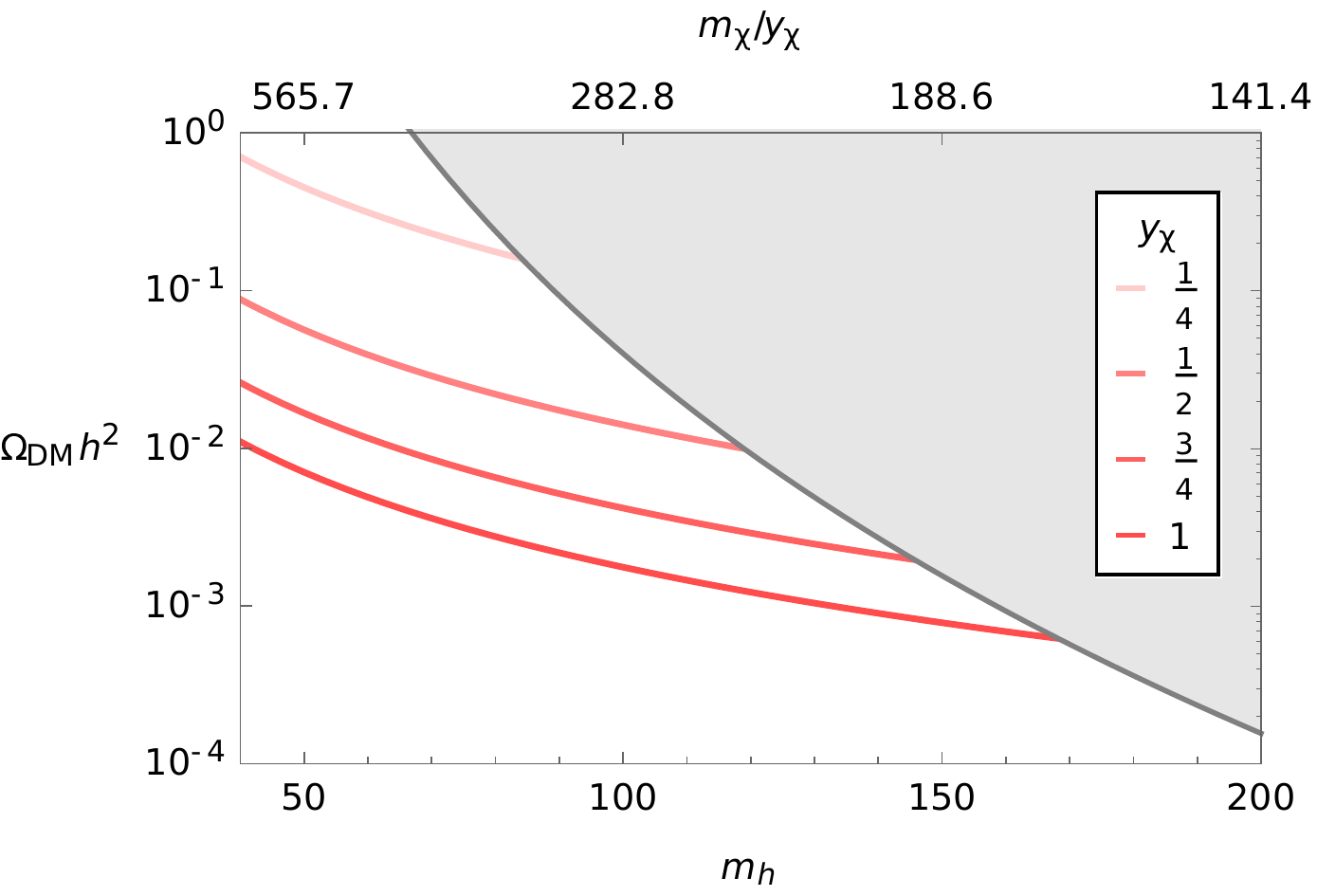}
    \caption{Example computation of the relic density for model \eqref{limiting234} with $\Lambda = 200$ GeV. In the gray region, $m_\chi < m_{h_D}$. Here we have considered the relic density of $N_f = 3/y_\chi$ fermions with Yukawa coupling $y_\chi$, such that this result can be compared with the results of the scan in the previous section.}
    \label{fig:relic}
\end{figure}

To illustrate the possible interplay between DM observations and the discovery of a new source of gravitational waves, we explore the region of correct relic abundance in the model \eqref{limiting234} with $\Lambda = 200$ GeV. The results are shown in Fig.~\ref{fig:relic}. 

These results are based on a toy model for DM, and many other scenarios could be considered. In particular, one could explore non-thermal production of DM and its relation with the scalar potential responsible for the PT. An alternative scenario has the heavy gauge bosons of the broken symmetry as the most important component of the dark matter relic abundance. Such a scenario was considered in \cite{Ko:2016fcd} for the symmetry breaking pattern $SU(3)/SU(2)$, and is sensitive to additional cosmological constraints from structure formation. 

\section{Discussion}
In this work we have considered the relic gravitational wave spectra from phase transitions in a hidden sector. These spectra can be related to the thermal parameters of the transition, which can be computed from first principles: $\beta/H$, the speed of the transition, $\Upsilon$, the latent heat, and $T_N$, nucleation temperature. We have distinguished between two limiting cases, with potentials \eqref{limiting234} and \eqref{limiting246}, which effectively capture the main classes of models. Furthermore, we have studied the effect of varying the quantum numbers of the scalar, the gauge coupling, and the number of coupled fermions. The results of these studies are summarized in Figs. \ref{fig:amps} and \ref{fig:freqs}, and some general conclusions are derived in section \ref{sec:results}. We find that although there is some degeneracy in the predictions, a level of model discrimination is possible. This is due to the fact that increasing the number of strongly coupled fermions, the rank of the group, or the number of scales involved all increase the visibility of a gravitational wave signal. Moreover the changes in thermal parameters due to each of these model changes are qualitatively different. In section \ref{sec:DMabundance}, we comment on the relic abundance of hidden sectors that could be constrained through their GW spectra.

A few caveats to our work. First, the renormalizable potential \eqref{234potential} does not have a zero-temperature potential barrier, such as could be the case for a singlet scalar with a cubic self-interaction. Phase transitions resulting from such a potential are qualitatively different, and the thermal corrections may restore the vacuum to a unique field value in such a way that no first order phase transition is expected. If a first order phase transition does occur, it may exhibit runaway behaviour, such that the GW spectrum from bubble shell collisions becomes relevant. This would lead to a different spectral shape, which in principle may be distinguishable in future experiments, for $T_N$ around the weak scale. A detailed analysis of such a scenario is beyond the scope of the present work.

Second, in the present work we have employed a high temperature expansion, Eq. \eqref{hightem}, which has a limited range of validity. Phase transitions not captured by this approximation may also give observable spectra; this is most noticeable in the results from the renormalizable potential \eqref{234potential}. In future work, it will be interesting to explore the models using the full thermal functions. Another possible extension is the inclusion of higher dimensional potential corrections to the two limiting potential forms considered here. An analysis with the inclusion of such operators will be presented in a future paper.

Finally, we have not calculated the wall velocities $v_w$ in the phase transitions, instead making conservative assumptions to calculate the spectra.
Calculating the bubble wall velocity for a general model with general parameters is a highly non-trivial task which we leave to future work. However, we can briefly comment on how measuring the bubble wall velocity can lead to further model discrimination. The wall velocity can be estimated in the limit that the departure from each particle in the plasma's equilibrium distributions are slowly varying near the bubble wall  \cite{Moore:1995si,John:2000zq,Kozaczuk:2014kva}. In this case the bubble wall velocity solves
\begin{equation}
    h_D ^{\prime \prime } -\frac{\partial V}{\partial h_D } = \eta v_w  \gamma h_D ^\prime h_D^2  
\end{equation}
for boundary conditions $h_D(-\infty) = 0$ and $h_D(\infty)=v (T_n)$, that is, the value of the non trivial minimum at the nucleation temperature. Only a particular value of the combined friction term on the right hand side will satisfy the boundary conditions, and since $\eta$ is determined by particle physics, the problem reduces to choosing an appropriate value for $v_w\gamma$ (where $\gamma$ is the Lorentz factor).
In the above $\eta$ can be written as a matrix product $G^T \Gamma ^{-1} F$ where $G$ and $F$ are vectors, whose components scale as $g^2$ or $y_\chi^2$. The matrix of coefficients scale either as $g^2 y_\chi^2$, $g^4$ or $y_\chi^4$. Therefore the bubble wall velocity can give more information on the size of both the gauge coupling and the fermion couplings, if present. 

Future work may also include further analysis of the internal hidden sector dynamics, including a thorough calculation of the thermal histories and relic abundances of hidden sector degrees of freedom. In this work we have chosen to focus on a decoupled hidden sector, but it is in principle straightforward to extend the results presented here to sectors with significant portal couplings. 

\vspace{0.3cm} {\bf Acknowledgments} 
The authors thank Anupam Mazumdar for useful discussions. DC also thanks Jose Miguel No and David Weir for useful discussions, and the Universidad Aut\'{o}noma de Madrid for hospitality during the completion of this paper. VS acknowledges support from the Science and Technology Facilities Council (ST/P000819/1).
TRIUMF receives federal funding via a contribution agreement with the National Research Council of Canada and the Natural Science and Engineering Research Council of Canada.

\section*{Appendix:  Thermal Corrections}
\subsection*{Renormalizable potential} 
To calculate thermal corrections we need to calculate the field dependent masses. For a dark Higgs sector the masses of the physical, goldstone, gauge boson and fermions are respectively
\begin{eqnarray}
m_H^2 &=& \partial _{h_D}^2 V = \Lambda ^4 \left( 3\frac{h_D^2}{v^4}-\frac{1}{v^2} \right) \\
m_G &=& \frac{1}{h_D} \partial _{h_D} V = \Lambda ^4 \left( \frac{h_D^2}{v^4}-\frac{1}{v^2} \right) \\
m_{GB} &=& \frac{gh_D}{2} \\
m_\chi &=& \frac{y_\chi h_D}{\sqrt{2}} \ .
\end{eqnarray}
We will work in the Landau gauge. The mass of the Goldstone mode is zero at the vev but becomes important when describing the phase transition. 
In the high temperature expansion the thermal corrections to the potential are
\begin{equation}
    V_T = \sum _{i \in {\rm bosons}} n_i \left( \frac{m^2}{24}T^2 - \frac{m^3}{12 \pi} T \right) +\sum _{i \in {\rm fermions}} n_i \frac{m^2}{48}T^2
\end{equation}
The next order in the expansion is given by a logarithm, which is cancelled by the zero temperature one-loop Coleman Weinberg potential. Note we have also ignored the constant term.
We find numerically that the high temperature expansion is valid almost exactly for $m^2<2\times T^2$. Our values of $n_i$ are 
\begin{eqnarray}
n_H=1 \quad n_{G}= 2N-1 \quad n_{GB}=3\times (2N-1) \quad n_f = 2 \times N \times N_f
\end{eqnarray}
where $N_f$ is the number of fermions and $N$ is the rank of the group. Note that we follow the standard practice of ignoring the second term in the high temperature expansion for Goldstones and Higgs $\sim m^{3}T$, such that the only cubic self-interaction comes from the gauge bosons. 

\subsection*{Non-renormalizable potential}
For the nonrenomalizable potential \eqref{246potential} we proceed as before, but here we assume the cubic corrections due to gauge bosons 
are subdominant compared to the zero temperature terms with alternating signs. 
This corresponds to only taking the first term in the high temperature expansion for every species.

\bibliographystyle{JHEP}
\bibliography{references}

\providecommand{\href}[2]{#2}\begingroup\raggedright\begin{thebibliography}{10}

\bibitem{Abbott:2016blz}
{\scshape Virgo, LIGO Scientific} collaboration, B.~P. Abbott et~al.,
  \emph{{Observation of Gravitational Waves from a Binary Black Hole Merger}},
  \href{http://dx.doi.org/10.1103/PhysRevLett.116.061102}{\emph{Phys. Rev.
  Lett.} {\bf 116} (2016) 061102}, [\href{http://arxiv.org/abs/1602.03837}{{\tt
  1602.03837}}].

\bibitem{Ellis:2017jgp}
J.~Ellis, A.~Hektor, G.~Hütsi, K.~Kannike, L.~Marzola, M.~Raidal et~al.,
  \emph{{Search for Dark Matter Effects on Gravitational Signals from Neutron
  Star Mergers}},
  \href{http://dx.doi.org/10.1016/j.physletb.2018.04.048}{\emph{Phys. Lett.}
  {\bf B781} (2018) 607--610}, [\href{http://arxiv.org/abs/1710.05540}{{\tt
  1710.05540}}].

\bibitem{Croon:2017zcu}
D.~Croon, A.~E. Nelson, C.~Sun, D.~G.~E. Walker and Z.-Z. Xianyu,
  \emph{{Hidden-Sector Spectroscopy with Gravitational Waves from Binary
  Neutron Stars}},
  \href{http://dx.doi.org/10.3847/2041-8213/aabe76}{\emph{Astrophys. J.} {\bf
  858} (2018) L2}, [\href{http://arxiv.org/abs/1711.02096}{{\tt 1711.02096}}].

\bibitem{Giudice:2016zpa}
G.~F. Giudice, M.~McCullough and A.~Urbano, \emph{{Hunting for Dark Particles
  with Gravitational Waves}},
  \href{http://dx.doi.org/10.1088/1475-7516/2016/10/001}{\emph{JCAP} {\bf 1610}
  (2016) 001}, [\href{http://arxiv.org/abs/1605.01209}{{\tt 1605.01209}}].

\bibitem{Palenzuela:2017kcg}
C.~Palenzuela, P.~Pani, M.~Bezares, V.~Cardoso, L.~Lehner and S.~Liebling,
  \emph{{Gravitational Wave Signatures of Highly Compact Boson Star Binaries}},
  \href{http://dx.doi.org/10.1103/PhysRevD.96.104058}{\emph{Phys. Rev.} {\bf
  D96} (2017) 104058}, [\href{http://arxiv.org/abs/1710.09432}{{\tt
  1710.09432}}].

\bibitem{Croon:2018ftb}
D.~Croon, M.~Gleiser, S.~Mohapatra and C.~Sun, \emph{{Gravitational Radiation
  Background from Boson Star Binaries}},
  \href{http://arxiv.org/abs/1802.08259}{{\tt 1802.08259}}.

\bibitem{White:2016nbo}
G.~A. White, \emph{{A Pedagogical Introduction to Electroweak Baryogenesis}}.
\newblock IOP Concise Physics. Morgan and Claypool, 2016,
  \href{http://dx.doi.org/10.1088/978-1-6817-4457-5}{10.1088/978-1-6817-4457-5}.

\bibitem{Morrissey:2012db}
D.~E. Morrissey and M.~J. Ramsey-Musolf, \emph{{Electroweak baryogenesis}},
  \href{http://dx.doi.org/10.1088/1367-2630/14/12/125003}{\emph{New J. Phys.}
  {\bf 14} (2012) 125003}, [\href{http://arxiv.org/abs/1206.2942}{{\tt
  1206.2942}}].

\bibitem{Trodden:1998ym}
M.~Trodden, \emph{{Electroweak baryogenesis}},
  \href{http://dx.doi.org/10.1103/RevModPhys.71.1463}{\emph{Rev. Mod. Phys.}
  {\bf 71} (1999) 1463--1500}, [\href{http://arxiv.org/abs/hep-ph/9803479}{{\tt
  hep-ph/9803479}}].

\bibitem{Dev:2016feu}
P.~S.~B. Dev and A.~Mazumdar, \emph{{Probing the Scale of New Physics by
  Advanced LIGO/VIRGO}},
  \href{http://dx.doi.org/10.1103/PhysRevD.93.104001}{\emph{Phys. Rev.} {\bf
  D93} (2016) 104001}, [\href{http://arxiv.org/abs/1602.04203}{{\tt
  1602.04203}}].

\bibitem{deVries:2017ncy}
J.~de~Vries, M.~Postma, J.~van~de Vis and G.~White, \emph{{Electroweak
  Baryogenesis and the Standard Model Effective Field Theory}},
  \href{http://dx.doi.org/10.1007/JHEP01(2018)089}{\emph{JHEP} {\bf 01} (2018)
  089}, [\href{http://arxiv.org/abs/1710.04061}{{\tt 1710.04061}}].

\bibitem{Balazs:2016yvi}
C.~Balazs, G.~White and J.~Yue, \emph{{Effective field theory, electric dipole
  moments and electroweak baryogenesis}},
  \href{http://dx.doi.org/10.1007/JHEP03(2017)030}{\emph{JHEP} {\bf 03} (2017)
  030}, [\href{http://arxiv.org/abs/1612.01270}{{\tt 1612.01270}}].

\bibitem{Akula:2017yfr}
S.~Akula, C.~Balázs, L.~Dunn and G.~White, \emph{{Electroweak baryogenesis in
  the $ {\mathbb{Z}}_3 $ -invariant NMSSM}},
  \href{http://dx.doi.org/10.1007/JHEP11(2017)051}{\emph{JHEP} {\bf 11} (2017)
  051}, [\href{http://arxiv.org/abs/1706.09898}{{\tt 1706.09898}}].

\bibitem{Balazs:2013cia}
C.~Balázs, A.~Mazumdar, E.~Pukartas and G.~White, \emph{{Baryogenesis, dark
  matter and inflation in the Next-to-Minimal Supersymmetric Standard Model}},
  \href{http://dx.doi.org/10.1007/JHEP01(2014)073}{\emph{JHEP} {\bf 01} (2014)
  073}, [\href{http://arxiv.org/abs/1309.5091}{{\tt 1309.5091}}].

\bibitem{Balazs:2004ae}
C.~Balazs, M.~Carena, A.~Menon, D.~E. Morrissey and C.~E.~M. Wagner, \emph{{The
  Supersymmetric origin of matter}},
  \href{http://dx.doi.org/10.1103/PhysRevD.71.075002}{\emph{Phys. Rev.} {\bf
  D71} (2005) 075002}, [\href{http://arxiv.org/abs/hep-ph/0412264}{{\tt
  hep-ph/0412264}}].

\bibitem{Lee:2004we}
C.~Lee, V.~Cirigliano and M.~J. Ramsey-Musolf, \emph{{Resonant relaxation in
  electroweak baryogenesis}},
  \href{http://dx.doi.org/10.1103/PhysRevD.71.075010}{\emph{Phys. Rev.} {\bf
  D71} (2005) 075010}, [\href{http://arxiv.org/abs/hep-ph/0412354}{{\tt
  hep-ph/0412354}}].

\bibitem{Balazs:2016tbi}
C.~Balazs, A.~Fowlie, A.~Mazumdar and G.~White, \emph{{Gravitational waves at
  aLIGO and vacuum stability with a scalar singlet extension of the Standard
  Model}}, \href{http://dx.doi.org/10.1103/PhysRevD.95.043505}{\emph{Phys.
  Rev.} {\bf D95} (2017) 043505}, [\href{http://arxiv.org/abs/1611.01617}{{\tt
  1611.01617}}].

\bibitem{Jinno:2017ixd}
R.~Jinno, S.~Lee, H.~Seong and M.~Takimoto, \emph{{Gravitational waves from
  first-order phase transitions: Towards model separation by bubble nucleation
  rate}}, \href{http://dx.doi.org/10.1088/1475-7516/2017/11/050}{\emph{JCAP}
  {\bf 1711} (2017) 050}, [\href{http://arxiv.org/abs/1708.01253}{{\tt
  1708.01253}}].

\bibitem{Matsui:2017ggm}
T.~Matsui, \emph{{Gravitational waves from the first order electroweak phase
  transition in the $Z_3$ symmetric singlet scalar model}},
  \href{http://dx.doi.org/10.1051/epjconf/201816805001}{\emph{EPJ Web Conf.}
  {\bf 168} (2018) 05001}, [\href{http://arxiv.org/abs/1709.05900}{{\tt
  1709.05900}}].

\bibitem{Huang:2017kzu}
F.~P. Huang and C.~S. Li, \emph{{Probing the baryogenesis and dark matter
  relaxed in phase transition by gravitational waves and colliders}},
  \href{http://dx.doi.org/10.1103/PhysRevD.96.095028}{\emph{Phys. Rev.} {\bf
  D96} (2017) 095028}, [\href{http://arxiv.org/abs/1709.09691}{{\tt
  1709.09691}}].

\bibitem{Baldes:2017ygu}
I.~Baldes, \emph{{Generation of Asymmetric Dark Matter and Gravitational
  Waves}},  in \emph{{29th Rencontres de Blois on Particle Physics and
  Cosmology Blois, France, May 28-June 2, 2017}}, 2017.
\newblock \href{http://arxiv.org/abs/1711.08251}{{\tt 1711.08251}}.

\bibitem{Demidov:2017lzf}
S.~V. Demidov, D.~S. Gorbunov and D.~V. Kirpichnikov, \emph{{Gravitational
  waves from phase transition in split NMSSM}},
  \href{http://dx.doi.org/10.1016/j.physletb.2018.02.007}{\emph{Phys. Lett.}
  {\bf B779} (2018) 191--194}, [\href{http://arxiv.org/abs/1712.00087}{{\tt
  1712.00087}}].

\bibitem{Chala:2018ari}
M.~Chala, C.~Krause and G.~Nardini, \emph{{Signals of the electroweak phase
  transition at colliders and gravitational wave observatories}},
  \href{http://arxiv.org/abs/1802.02168}{{\tt 1802.02168}}.

\bibitem{Hashino:2018zsi}
K.~Hashino, M.~Kakizaki, S.~Kanemura, P.~Ko and T.~Matsui, \emph{{Gravitational
  waves from first order electroweak phase transition in models with the
  $U(1)_X^{}$ gauge symmetry}},  \href{http://arxiv.org/abs/1802.02947}{{\tt
  1802.02947}}.

\bibitem{Grojean:2006bp}
C.~Grojean and G.~Servant, \emph{{Gravitational Waves from Phase Transitions at
  the Electroweak Scale and Beyond}},
  \href{http://dx.doi.org/10.1103/PhysRevD.75.043507}{\emph{Phys. Rev.} {\bf
  D75} (2007) 043507}, [\href{http://arxiv.org/abs/hep-ph/0607107}{{\tt
  hep-ph/0607107}}].

\bibitem{Caprini:2015zlo}
C.~Caprini et~al., \emph{{Science with the space-based interferometer eLISA.
  II: Gravitational waves from cosmological phase transitions}},
  \href{http://dx.doi.org/10.1088/1475-7516/2016/04/001}{\emph{JCAP} {\bf 1604}
  (2016) 001}, [\href{http://arxiv.org/abs/1512.06239}{{\tt 1512.06239}}].

\bibitem{Apreda:2001us}
R.~Apreda, M.~Maggiore, A.~Nicolis and A.~Riotto, \emph{{Gravitational waves
  from electroweak phase transitions}},
  \href{http://dx.doi.org/10.1016/S0550-3213(02)00264-X}{\emph{Nucl. Phys.}
  {\bf B631} (2002) 342--368}, [\href{http://arxiv.org/abs/gr-qc/0107033}{{\tt
  gr-qc/0107033}}].

\bibitem{Hindmarsh:2016lnk}
M.~Hindmarsh, \emph{{Sound shell model for acoustic gravitational wave
  production at a first-order phase transition in the early Universe}},
  \href{http://dx.doi.org/10.1103/PhysRevLett.120.071301}{\emph{Phys. Rev.
  Lett.} {\bf 120} (2018) 071301}, [\href{http://arxiv.org/abs/1608.04735}{{\tt
  1608.04735}}].

\bibitem{Hindmarsh:2017gnf}
M.~Hindmarsh, S.~J. Huber, K.~Rummukainen and D.~J. Weir, \emph{{Shape of the
  acoustic gravitational wave power spectrum from a first order phase
  transition}}, \href{http://dx.doi.org/10.1103/PhysRevD.96.103520}{\emph{Phys.
  Rev.} {\bf D96} (2017) 103520}, [\href{http://arxiv.org/abs/1704.05871}{{\tt
  1704.05871}}].

\bibitem{Cutting:2018tjt}
D.~Cutting, M.~Hindmarsh and D.~J. Weir, \emph{{Gravitational waves from vacuum
  first-order phase transitions: from the envelope to the lattice}},
  \href{http://arxiv.org/abs/1802.05712}{{\tt 1802.05712}}.

\bibitem{Croon:2018new}
D.~Croon and G.~White, \emph{{Exotic Gravitational Wave Signatures from
  Simultaneous Phase Transitions}},
  \href{http://arxiv.org/abs/1803.05438}{{\tt 1803.05438}}.

\bibitem{Schwaller:2015tja}
P.~Schwaller, \emph{{Gravitational Waves from a Dark Phase Transition}},
  \href{http://dx.doi.org/10.1103/PhysRevLett.115.181101}{\emph{Phys. Rev.
  Lett.} {\bf 115} (2015) 181101}, [\href{http://arxiv.org/abs/1504.07263}{{\tt
  1504.07263}}].

\bibitem{Bodeker:2017cim}
D.~Bodeker and G.~D. Moore, \emph{{Electroweak Bubble Wall Speed Limit}},
  \href{http://dx.doi.org/10.1088/1475-7516/2017/05/025}{\emph{JCAP} {\bf 1705}
  (2017) 025}, [\href{http://arxiv.org/abs/1703.08215}{{\tt 1703.08215}}].

\bibitem{Croon:2015naa}
D.~Croon, V.~Sanz and E.~R.~M. Tarrant, \emph{{Reheating with a composite Higgs
  boson}}, \href{http://dx.doi.org/10.1103/PhysRevD.94.045010}{\emph{Phys.
  Rev.} {\bf D94} (2016) 045010}, [\href{http://arxiv.org/abs/1507.04653}{{\tt
  1507.04653}}].

\bibitem{Croon:2015fza}
D.~Croon, V.~Sanz and J.~Setford, \emph{{Goldstone Inflation}},
  \href{http://dx.doi.org/10.1007/JHEP10(2015)020}{\emph{JHEP} {\bf 10} (2015)
  020}, [\href{http://arxiv.org/abs/1503.08097}{{\tt 1503.08097}}].

\bibitem{Geller:2018mwu}
M.~Geller, A.~Hook, R.~Sundrum and Y.~Tsai, \emph{{Primordial Anisotropies in
  the Gravitational Wave Background from Cosmological Phase Transitions}},
  \href{http://arxiv.org/abs/1803.10780}{{\tt 1803.10780}}.

\bibitem{Quiros:1999jp}
M.~Quiros, \emph{{Finite temperature field theory and phase transitions}},  in
  \emph{{Proceedings, Summer School in High-energy physics and cosmology:
  Trieste, Italy, June 29-July 17, 1998}}, pp.~187--259, 1999.
\newblock \href{http://arxiv.org/abs/hep-ph/9901312}{{\tt hep-ph/9901312}}.

\bibitem{Gorda:2018hvi}
T.~Gorda, A.~Helset, L.~Niemi, T.~V.~I. Tenkanen and D.~J. Weir,
  \emph{{Electroweak phase transition and dimensional reduction of the
  Two-Higgs-Doublet Model}},  \href{http://arxiv.org/abs/1802.05056}{{\tt
  1802.05056}}.

\bibitem{Curtin:2016urg}
D.~Curtin, P.~Meade and H.~Ramani, \emph{{Thermal Resummation and Phase
  Transitions}},  \href{http://arxiv.org/abs/1612.00466}{{\tt 1612.00466}}.

\bibitem{Akula:2016gpl}
S.~Akula, C.~Balázs and G.~A. White, \emph{{Semi-analytic techniques for
  calculating bubble wall profiles}},
  \href{http://dx.doi.org/10.1140/epjc/s10052-016-4519-5}{\emph{Eur. Phys. J.}
  {\bf C76} (2016) 681}, [\href{http://arxiv.org/abs/1608.00008}{{\tt
  1608.00008}}].

\bibitem{King:2014iia}
S.~F. King, \emph{{A to Z of Flavour with Pati-Salam}},
  \href{http://dx.doi.org/10.1007/JHEP08(2014)130}{\emph{JHEP} {\bf 08} (2014)
  130}, [\href{http://arxiv.org/abs/1406.7005}{{\tt 1406.7005}}].

\bibitem{Ramsey-Musolf:2017tgh}
M.~J. Ramsey-Musolf, P.~Winslow and G.~White, \emph{{Color Breaking
  Baryogenesis}},
  \href{http://dx.doi.org/10.1103/PhysRevD.97.123509}{\emph{Phys. Rev.} {\bf
  D97} (2018) 123509}, [\href{http://arxiv.org/abs/1708.07511}{{\tt
  1708.07511}}].

\bibitem{Patel:2013zla}
H.~H. Patel, M.~J. Ramsey-Musolf and M.~B. Wise, \emph{{Color Breaking in the
  Early Universe}},
  \href{http://dx.doi.org/10.1103/PhysRevD.88.015003}{\emph{Phys. Rev.} {\bf
  D88} (2013) 015003}, [\href{http://arxiv.org/abs/1303.1140}{{\tt
  1303.1140}}].

\bibitem{Kaplan:2009de}
D.~E. Kaplan, G.~Z. Krnjaic, K.~R. Rehermann and C.~M. Wells, \emph{{Atomic
  Dark Matter}},
  \href{http://dx.doi.org/10.1088/1475-7516/2010/05/021}{\emph{JCAP} {\bf 1005}
  (2010) 021}, [\href{http://arxiv.org/abs/0909.0753}{{\tt 0909.0753}}].

\bibitem{Petraki:2011mv}
K.~Petraki, M.~Trodden and R.~R. Volkas, \emph{{Visible and dark matter from a
  first-order phase transition in a baryon-symmetric universe}},
  \href{http://dx.doi.org/10.1088/1475-7516/2012/02/044}{\emph{JCAP} {\bf 1202}
  (2012) 044}, [\href{http://arxiv.org/abs/1111.4786}{{\tt 1111.4786}}].

\bibitem{Ma:2017vzm}
Y.~Wu, T.~Ma, B.~Zhang and G.~Cacciapaglia, \emph{{Composite Dark Matter and
  Higgs}}, \href{http://dx.doi.org/10.1007/JHEP11(2017)058}{\emph{JHEP} {\bf
  11} (2017) 058}, [\href{http://arxiv.org/abs/1703.06903}{{\tt 1703.06903}}].

\bibitem{Ackerman:mha}
L.~Ackerman, M.~R. Buckley, S.~M. Carroll and M.~Kamionkowski, \emph{{Dark
  Matter and Dark Radiation}},
  \href{http://dx.doi.org/10.1103/PhysRevD.79.023519,
  10.1142/9789814293792_0021}{\emph{Phys. Rev.} {\bf D79} (2009) 023519},
  [\href{http://arxiv.org/abs/0810.5126}{{\tt 0810.5126}}].

\bibitem{Agrawal:2017rvu}
P.~Agrawal, F.-Y. Cyr-Racine, L.~Randall and J.~Scholtz, \emph{{Dark
  Catalysis}},
  \href{http://dx.doi.org/10.1088/1475-7516/2017/08/021}{\emph{JCAP} {\bf 1708}
  (2017) 021}, [\href{http://arxiv.org/abs/1702.05482}{{\tt 1702.05482}}].

\bibitem{Buckley:2017ttd}
M.~R. Buckley and A.~DiFranzo, \emph{{Collapsed Dark Matter Structures}},
  \href{http://dx.doi.org/10.1103/PhysRevLett.120.051102}{\emph{Phys. Rev.
  Lett.} {\bf 120} (2018) 051102}, [\href{http://arxiv.org/abs/1707.03829}{{\tt
  1707.03829}}].

\bibitem{Krnjaic:2014xza}
G.~Krnjaic and K.~Sigurdson, \emph{{Big Bang Darkleosynthesis}},
  \href{http://dx.doi.org/10.1016/j.physletb.2015.11.001}{\emph{Phys. Lett.}
  {\bf B751} (2015) 464--468}, [\href{http://arxiv.org/abs/1406.1171}{{\tt
  1406.1171}}].

\bibitem{Coleman:1977py}
S.~R. Coleman, \emph{{The Fate of the False Vacuum. 1. Semiclassical Theory}},
  \href{http://dx.doi.org/10.1103/PhysRevD.15.2929,
  10.1103/PhysRevD.16.1248}{\emph{Phys. Rev.} {\bf D15} (1977) 2929--2936}.

\bibitem{Wainwright:2011kj}
C.~L. Wainwright, \emph{{CosmoTransitions: Computing Cosmological Phase
  Transition Temperatures and Bubble Profiles with Multiple Fields}},
  \href{http://dx.doi.org/10.1016/j.cpc.2012.04.004}{\emph{Comput. Phys.
  Commun.} {\bf 183} (2012) 2006--2013},
  [\href{http://arxiv.org/abs/1109.4189}{{\tt 1109.4189}}].

\bibitem{Weir:2017wfa}
D.~J. Weir, \emph{{Gravitational waves from a first order electroweak phase
  transition: a brief review}},
  \href{http://dx.doi.org/10.1098/rsta.2017.0126}{\emph{Phil. Trans. Roy. Soc.
  Lond.} {\bf A376} (2018) 20170126},
  [\href{http://arxiv.org/abs/1705.01783}{{\tt 1705.01783}}].

\bibitem{Espinosa:2010hh}
J.~R. Espinosa, T.~Konstandin, J.~M. No and G.~Servant, \emph{{Energy Budget of
  Cosmological First-order Phase Transitions}},
  \href{http://dx.doi.org/10.1088/1475-7516/2010/06/028}{\emph{JCAP} {\bf 1006}
  (2010) 028}, [\href{http://arxiv.org/abs/1004.4187}{{\tt 1004.4187}}].

\bibitem{Huber:2007vva}
S.~J. Huber and T.~Konstandin, \emph{{Production of gravitational waves in the
  nMSSM}}, \href{http://dx.doi.org/10.1088/1475-7516/2008/05/017}{\emph{JCAP}
  {\bf 0805} (2008) 017}, [\href{http://arxiv.org/abs/0709.2091}{{\tt
  0709.2091}}].

\bibitem{Delaunay:2007wb}
C.~Delaunay, C.~Grojean and J.~D. Wells, \emph{{Dynamics of Non-renormalizable
  Electroweak Symmetry Breaking}},
  \href{http://dx.doi.org/10.1088/1126-6708/2008/04/029}{\emph{JHEP} {\bf 04}
  (2008) 029}, [\href{http://arxiv.org/abs/0711.2511}{{\tt 0711.2511}}].

\bibitem{Vaskonen:2016yiu}
V.~Vaskonen, \emph{{Electroweak baryogenesis and gravitational waves from a
  real scalar singlet}},
  \href{http://dx.doi.org/10.1103/PhysRevD.95.123515}{\emph{Phys. Rev.} {\bf
  D95} (2017) 123515}, [\href{http://arxiv.org/abs/1611.02073}{{\tt
  1611.02073}}].

\bibitem{Bodeker:2009qy}
D.~Bodeker and G.~D. Moore, \emph{{Can electroweak bubble walls run away?}},
  \href{http://dx.doi.org/10.1088/1475-7516/2009/05/009}{\emph{JCAP} {\bf 0905}
  (2009) 009}, [\href{http://arxiv.org/abs/0903.4099}{{\tt 0903.4099}}].

\bibitem{Moore:2014lga}
C.~J. Moore, R.~H. Cole and C.~P.~L. Berry, \emph{{Gravitational-wave
  sensitivity curves}},
  \href{http://dx.doi.org/10.1088/0264-9381/32/1/015014}{\emph{Class. Quant.
  Grav.} {\bf 32} (2015) 015014}, [\href{http://arxiv.org/abs/1408.0740}{{\tt
  1408.0740}}].

\bibitem{Thrane:2013oya}
E.~Thrane and J.~D. Romano, \emph{{Sensitivity curves for searches for
  gravitational-wave backgrounds}},
  \href{http://dx.doi.org/10.1103/PhysRevD.88.124032}{\emph{Phys. Rev.} {\bf
  D88} (2013) 124032}, [\href{http://arxiv.org/abs/1310.5300}{{\tt
  1310.5300}}].

\bibitem{Chacko:2015noa}
Z.~Chacko, Y.~Cui, S.~Hong and T.~Okui, \emph{{Hidden dark matter sector, dark
  radiation, and the CMB}},
  \href{http://dx.doi.org/10.1103/PhysRevD.92.055033}{\emph{Phys. Rev.} {\bf
  D92} (2015) 055033}, [\href{http://arxiv.org/abs/1505.04192}{{\tt
  1505.04192}}].

\bibitem{Krnjaic:2015mbs}
G.~Krnjaic, \emph{{Probing Light Thermal Dark-Matter With a Higgs Portal
  Mediator}}, \href{http://dx.doi.org/10.1103/PhysRevD.94.073009}{\emph{Phys.
  Rev.} {\bf D94} (2016) 073009}, [\href{http://arxiv.org/abs/1512.04119}{{\tt
  1512.04119}}].

\bibitem{Lee:2013bua}
H.~M. Lee, M.~Park and V.~Sanz, \emph{{Gravity-mediated (or Composite) Dark
  Matter}}, \href{http://dx.doi.org/10.1140/epjc/s10052-014-2715-8}{\emph{Eur.
  Phys. J.} {\bf C74} (2014) 2715}, [\href{http://arxiv.org/abs/1306.4107}{{\tt
  1306.4107}}].

\bibitem{Ko:2016fcd}
P.~Ko and Y.~Tang, \emph{{Residual Non-Abelian Dark Matter and Dark
  Radiation}},
  \href{http://dx.doi.org/10.1016/j.physletb.2017.02.033}{\emph{Phys. Lett.}
  {\bf B768} (2017) 12--17}, [\href{http://arxiv.org/abs/1609.02307}{{\tt
  1609.02307}}].

\bibitem{Moore:1995si}
G.~D. Moore and T.~Prokopec, \emph{{How fast can the wall move? A Study of the
  electroweak phase transition dynamics}},
  \href{http://dx.doi.org/10.1103/PhysRevD.52.7182}{\emph{Phys. Rev.} {\bf D52}
  (1995) 7182--7204}, [\href{http://arxiv.org/abs/hep-ph/9506475}{{\tt
  hep-ph/9506475}}].

\bibitem{John:2000zq}
P.~John and M.~G. Schmidt, \emph{{Do stops slow down electroweak bubble
  walls?}}, \href{http://dx.doi.org/10.1016/S0550-3213(00)00768-9,
  10.1016/S0550-3213(02)01014-3}{\emph{Nucl. Phys.} {\bf B598} (2001)
  291--305}, [\href{http://arxiv.org/abs/hep-ph/0002050}{{\tt
  hep-ph/0002050}}].

\bibitem{Kozaczuk:2014kva}
J.~Kozaczuk, S.~Profumo, L.~S. Haskins and C.~L. Wainwright,
  \emph{{Cosmological Phase Transitions and their Properties in the NMSSM}},
  \href{http://dx.doi.org/10.1007/JHEP01(2015)144}{\emph{JHEP} {\bf 01} (2015)
  144}, [\href{http://arxiv.org/abs/1407.4134}{{\tt 1407.4134}}].

\end{thebibliography}\endgroup

\end{document}